\documentclass[graybox,envcountchap,sectrefs,natbib]{svmono}
\usepackage{mathptmx} 
\usepackage{helvet}       
\usepackage{courier}    
\usepackage{type1cm}  
\usepackage{amsfonts}
\usepackage{amssymb}
\usepackage{xspace}
\usepackage{chapterbib}
\usepackage{makeidx}         
\usepackage{multicol}        
\usepackage[bottom]{footmisc}
\usepackage{graphicx}        

\begin{document}
\mainmatter
\setcounter{page}{0} 
\setcounter{chapter}{1} %

%
%
\let\jnl=\rmfamily
\def\refe@jnl#1{{\jnl#1}}%

\newcommand\aj{\refe@jnl{AJ}}%
\newcommand\actaa{\refe@jnl{Acta Astron.}}%
\newcommand\araa{\refe@jnl{ARA\&A}}%
\newcommand\apj{\refe@jnl{ApJ}}%
\newcommand\apjl{\refe@jnl{ApJ}}%
\newcommand\apjs{\refe@jnl{ApJS}}%
\newcommand\ao{\refe@jnl{Appl.~Opt.}}%
\newcommand\apss{\refe@jnl{Ap\&SS}}%
\newcommand\aap{\refe@jnl{A\&A}}%
\newcommand\aapr{\refe@jnl{A\&A~Rev.}}%
\newcommand\aaps{\refe@jnl{A\&AS}}%
\newcommand\azh{\refe@jnl{AZh}}%
\newcommand\gca{\refe@jnl{GeoCh.Act}}%
\newcommand\grl{\refe@jnl{Geo.Res.Lett.}}%
\newcommand\jgr{\refe@jnl{J.Geoph.Res.}}%
\newcommand\memras{\refe@jnl{MmRAS}}%
\newcommand\jrasc{\refe@jnl{J.RoySocCan}}%
\newcommand\mnras{\refe@jnl{MNRAS}}%
\newcommand\na{\refe@jnl{New A}}%
\newcommand\nar{\refe@jnl{New A Rev.}}%
\newcommand\pra{\refe@jnl{Phys.~Rev.~A}}%
\newcommand\prb{\refe@jnl{Phys.~Rev.~B}}%
\newcommand\prc{\refe@jnl{Phys.~Rev.~C}}%
\newcommand\prd{\refe@jnl{Phys.~Rev.~D}}%
\newcommand\pre{\refe@jnl{Phys.~Rev.~E}}%
\newcommand\prl{\refe@jnl{Phys.~Rev.~Lett.}}%
\newcommand\pasa{\refe@jnl{PASA}}%
\newcommand\pasp{\refe@jnl{PASP}}%
\newcommand\pasj{\refe@jnl{PASJ}}%
\newcommand\skytel{\refe@jnl{S\&T}}%
\newcommand\solphys{\refe@jnl{Sol.~Phys.}}%
\newcommand\sovast{\refe@jnl{Soviet~Ast.}}%
\newcommand\ssr{\refe@jnl{Space~Sci.~Rev.}}%
\newcommand\nat{\refe@jnl{Nature}}%
\newcommand\iaucirc{\refe@jnl{IAU~Circ.}}%
\newcommand\aplett{\refe@jnl{Astrophys.~Lett.}}%
\newcommand\apspr{\refe@jnl{Astrophys.~Space~Phys.~Res.}}%
\newcommand\nphysa{\refe@jnl{Nucl.~Phys.~A}}%
\newcommand\physrep{\refe@jnl{Phys.~Rep.}}%
\newcommand\procspie{\refe@jnl{Proc.~SPIE}}%

\newcommand{\Al}{$^{26}$Al\xspace}
\newcommand{\al}{$^{26}$Al\xspace}
\newcommand{\Be}{$^{7}$Be\xspace}
\newcommand{\be}{$^{7}$Be\xspace}
\newcommand{\bem}{$^{10}$Be\xspace}
\newcommand{\ca}{$^{44}$Ca\xspace}
\newcommand{\Ca}{$^{44}$Ca\xspace}
\newcommand{\cam}{$^{41}$Ca\xspace}
\newcommand{\Co}{$^{56}$Co\xspace}
\newcommand{\co}{$^{56}$Co\xspace}
\newcommand{\csm}{$^{135}$Cs\xspace}
\newcommand{\ct}{$^{13}$C\xspace}
\newcommand{\ci}{$^{57}$Co\xspace}
\newcommand{\Ci}{$^{57}$Co\xspace}
\newcommand{\ch}{$^{60}$Co\xspace}
\newcommand{\Ch}{$^{60}$Co\xspace}
\newcommand{\Cl}{$^{36}$Cl\xspace}
\newcommand{\li}{$^{7}$Li\xspace}
\newcommand{\Li}{$^{7}$Li\xspace}
\newcommand{\Fe}{$^{60}$Fe\xspace}
\newcommand{\fh}{$^{60}$Fe\xspace}
\newcommand{\fe}{$^{56}$Fe\xspace}
\newcommand{\Fr}{$^{57}$Fe\xspace}
\newcommand{\fr}{$^{57}$Fe\xspace}
\newcommand{\mg}{$^{26}$Mg\xspace}
\newcommand{\Mg}{$^{26}$Mg\xspace}
\newcommand{\mn}{$^{54}$Mn\xspace}
\newcommand{\Na}{$^{22}$Na\xspace}
\newcommand{\Ne}{$^{22}$Ne\xspace}
\newcommand{\Ni}{$^{56}$Ni\xspace}
\newcommand{\nh}{$^{60}$Ni\xspace}
\newcommand{\Nh}{$^{60}$Ni\xspace}
\newcommand\nuk[2]{$\rm ^{\rm #2} #1$}  
\newcommand{\pd}{$^{107}$Pd\xspace}
\newcommand{\pb}{$^{205}$Pb\xspace}
\newcommand{\tc}{$^{99}$Tc\xspace}
\newcommand{\Sc}{$^{44}$Sc\xspace}
\newcommand{\Ti}{$^{44}$Ti\xspace}
\newcommand{\ti}{$^{44}$Ti\xspace}
\def\aa{$\alpha$}
\newcommand{\about}{$\simeq$}
\newcommand{\cms}{cm\ensuremath{^{-2}} s\ensuremath{^{-1}}\xspace}
\newcommand{\degree}{$^{\circ}$}
\newcommand{\flux}{ph~cm\ensuremath{^{-2}} s\ensuremath{^{-1}}\xspace}
\newcommand{\fluxrad}{ph~cm$^{-2}$s$^{-1}$rad$^{-1}$\ }
\newcommand{\ga}{\ensuremath{\gamma}}
\newcommand{\gam}{\ensuremath{\gamma}}
\def\nn{$\nu$}
\def\ra{$\rightarrow$}
\newcommand{\Msol}{M\ensuremath{_\odot}\xspace}
\newcommand{\msol}{M\ensuremath{_\odot}\xspace}
\newcommand{\Msolppc}{M\ensuremath{_\odot} pc\ensuremath{^{-2}}{\xspace}}
\newcommand{\Msolpy}{M\ensuremath{_\odot} y\ensuremath{^{-1}}{\xspace}}
\newcommand{\msb}{M\ensuremath{_\odot}\xspace}
\newcommand{\Msun}{M\ensuremath{_\odot}\xspace}
\newcommand{\Rsun}{R\ensuremath{_\odot}\xspace}
\newcommand{\rsun}{R\ensuremath{_\odot}\xspace}
\newcommand{\Lsun}{L\ensuremath{_\odot}\xspace}
\newcommand{\lsun}{L\ensuremath{_\odot}\xspace}
\newcommand{\solar}{\ensuremath{_\odot}\xspace}
\newcommand{\zs}{Z\ensuremath{_\odot}\xspace}




%
%
\chapauthor{Donald D. Clayton\footnote{Clemson University, Clemson SC 29634-0978, USA}}
\chapter{The Role of Radioactive Isotopes in Astrophysics}
\label{Ch:radioactivity}

\section{History of Nucleosynthesis and Radioactivity}
\label{sec:2-1}
\par\indent Radioactive nuclei can not be not infinitely old or they would now be gone. This point had first been made \index{Rutherford, E.}  by Lord Rutherford, who concluded in 1929 \citep{Ruth29} that the elements had been created in the sun 100~Myr ago and had somehow got from sun to earth! Terrible astrophysics, but his argument showed profound appreciation of the conundrum of live radioactivity on earth; namely, the nuclei must be created. By the middle of the 20$^{\rm{th}}$ century, the consequences had progressed no further. The question, ``Why aren't they all gone?'' went largely unspoken.

	Radioactivity \index{radioactivity} had been discovered by Henri Becquerel \index{Becquerel, A.H.} in 1896 when he showed that the element uranium emitted radiation that would develop images of the uranium salts on an adjacent photographic plate. Marie Curie \index{Curie} coined the term \emph{radioactivity} \index{radioactivity} to describe this phenomenon after she demonstrated that other elements possessed the same spontaneous property. Rutherford separated the rays into three types -- $\alpha$, $\beta$, and $\gamma$ -- by their physical properties.  These are described in Chapter~1.

	In mid-$20^{\rm{th}}$ century, astronomers discovered metal-poor stars in increasing numbers, especially after the introduction of the CCD chip, which greatly increased the sensitivity of spectroscopic observations of stars. The abundances relative to hydrogen in old low-mass stars reveals the initial composition of the gas from which those stars formed. Observed metallicities ranged over a continuum of values from (Fe/H)/(Fe/H)$_{\odot}$ = 10$^{-4}$ to 1 and even greater, so the synthesis of metals must have occurred after the \index{big bang} Big Bang. Today we know that creation of the chemical elements occurs in stars by synthesis of them from initial H and He. By the same token, natural radioactivity must be the product of nucleosynthesis in stars.

\subsection{Two very different Pioneers}
\label{sec:2-1-1}
	There could hardly be two more different scientists than the two that made the first significant steps toward nucleosynthesis in stars of the elements heavier than carbon. Both were Europeans disrupted in careers by world war II. The originator of nucleosynthesis in stars arose from non-privileged Yorkshire mill villages. The pioneer in glimpsing how nuclear magic numbers can be observed in isotopic abundances arose from a distinguished Viennese family of geologists.

\subsubsection*{Sir Fred Hoyle (1946, 1954)}
	In a flash of astrophysical brilliance, Fred Hoyle \index{Hoyle, F.}  created the theory of nucleosynthesis \index{nucleosynthesis}  in stars with two monumental papers \citep{Hoyle46, Hoyle54}. The first \citep{Hoyle46} demonstrated that stars will naturally evolve to central temperatures 4$\times 10^9$ K and greater, and that nuclear statistical equilibrium (NSE) at high temperature $T$ creates an abundance peak at $^{56}$Fe that Hoyle interpreted as the explanation of that observed abundance peak. This explains why a graph of the abundances of chemical elements in the sun as a function of atomic weight declines almost monotonically from H to Sc (A=1 to 45), whereafter it increases sharply to a broad abundance peak centered on A=56 (Fe).  Readers unfamiliar with the meaning and issues of abundances in the universe can find an eloquent essay in the introductory chapter of \citet{Clay03}.

Hoyle's second paper \citep{Hoyle54} described how the elements from carbon to nickel are synthesized during the advanced evolution of massive stars. This is called \emph{primary nucleosynthesis} because the elements are fused from the initial H and He of the massive stars. Hoyle described how the \emph{ashes} of each thermonuclear burning epoch became, upon contraction and heating, the fuel for the next group of elements synthesized. When its core collapses, much of the overlying new elements is returned to the interstellar medium.  Such nucleosynthesis creates new heavy nuclei and is responsible for increasing the metallicity of the interstellar medium \index{chemical evolution!cosmic} as it ages.

Because the origin of our chemical elements is one of the grand theories of science, \citep{Clay07}went to some lengths to create what he called {\it Hoyle's equation} as determined from careful reading of his 1954 paper \citep{Hoyle54}. Hoyle's words and calculations pointed clearly to ideas of nucleosynthesis in stars that he was advancing for the first time and that are more sweeping than detail-oriented sequels. Hoyle's discussion was phrased in terms of the mass $\Delta m_{new}$ of new primary isotopes that are being ejected from massive stars.
 His basic approach to stellar nucleosynthesis is

\begin{equation}
\frac{dm_{new}}{dt} =  H_{nucl}
\end{equation}

\noindent where 
\begin{equation}
H_{nucl} = B(t')\textbf{Ev}(t'-t) \Sigma_{k}\Delta m_{k}
\end{equation}
 is the \emph{Hoyle nucleosynthesis rate}.  B$(t')$ is the stellar birthrate of stars having total mass such that they evolve to end their lives at time $t$, $\textbf{Ev}(t',t)$ is an operator (rather than a number) that expresses the nuclear and stellar evolution during its lifetime from $t'$ to $t$, and $\Sigma m_{\rm{k}}$ is the mass of isotope $k$ ejected at time $t$. Then a sum over all presolar birthdates $t'$ selects the appropriate stellar masses. Stellar evolution was only dimly perceived in 1953 when Hoyle wrote this paper. The structure of red giants was its current literature frontier, and Hoyle was perhaps the world's leading expert on its ideas, enabling him to discern the more fruitful subsequent evolution that he advanced in this 1954 paper.

 	Hoyle subtitled this paper \emph{The synthesis of elements from Carbon to Nickel}.  These primary isotopes are ejected from massive stars that progress through a series of core evolutions that Hoyle laid out for the first time. Hoyle explained that gravitational contraction causes temperature increases after each nuclear fuel is consumed, and he described the nuclear burning during each advanced core evolution. Because those massive stars all evolve almost instantaneously in comparison with galactic timescale, Hoyle takes $B_{\rm{M}>}(t)$ to be the birthrate of all such massive stars at time $t$, and it clearly equals their death rate at the same time if the numbers of stars are to change only slowly. The subscript $M>$ characterizes stars too massive to become white dwarfs. For those large-mass stars  \citet{Hoyle54} predicted that collapse of the final central evolved core is inevitable. For those massive stars Hoyle's equation expresses the rate of ejection of new primary isotopes from C to Ni as

\begin{equation}
\frac{dm(C-Ni)}{dt} =B_{M>}(t) \textbf{Ev}_{nucl} \Sigma_{k}\Delta m_k
\end{equation}

For these short-lived stars, $\textbf{Ev}$ is an operator (rather than a number) that expresses the nuclear and stellar evolution during the stellar lifetime. It replaces the initial composition of the star by the composition it has attained at the time when its core collapses.  Hoyle attributed the mass and identity $k$ of new primary isotopes ejected per massive star to the following successive core burning phases: $^{12}$C \index{isotopes!12C} and $^{16}$O from \index{isotopes!20Ne} He burning; $^{20}$Ne, $^{23}$Na, and $^{24}$Mg from C burning; additional $^{16}$O and $^{24}$Mg \index{isotopes!24Mg} \index{isotopes!16O} from Ne burning; $^{28}$Si \index{isotopes!28Si} and $^{32}$S from O burning; $^{32}$S, $^{36}$Ar and $^{40}$Ca \index{isotopes!40Ca} from photoalpha reactions on $^{32}$S \index{isotopes!32S} and heavier alpha nuclei during later heating of O-burned matter by the inevitable contraction; and finally\index{isotopes!52Cr}\index{isotopes!56Fe} \index{isotopes!60Ni}    $^{52}$Cr, $^{56}$Fe, $^{60}$Ni from subsequent \index{process!nuclear statistical equilibrium} nuclear statistical equilibrium. Hoyle also correctly stated that \index{neutrino} neutrino emission governs the collapse timescale when core temperature exceeds T=3$\times$10$^9$ K.

Hoyle's equation expresses a breathtakingly modern view of the metallicity-increasing nucleosynthesis during galactic history. Hoyle missed only the complete photonuclear rearrangement during Si burning and the $n/p$ ratio within the NSE. But his equation, given above, remains correct today. Like Schroedinger's equation, for which much work is required to determine the appropriate Hamiltonian operator used within it, so Hoyle's equation involves much work to determine the Hoyle nuclear evolution $H_{nucl}$.

\citet{Hoyle54} also distinguished the idea of secondary nuclei, those whose created abundance derives from initial seed concentrations of primary nuclei that had been created in previous massive stars, thereby seeding the interstellar matter with primary elements. These seed nuclei are required in order that secondary nuclei be produced from them in stars. He emphasized especially $^{14}$N, $^{18}$O, $^{19}$F and $^{22}$Ne in that regard, each of which depends on initial primary C and O nuclei within the initial composition of later-generation massive stars.  Their yields of secondary nuclei do not obey Hoyle's equation but instead are proportional to the initial metallicity of each star. Hoyle's paper also first noted that $^{22}$Ne would be a source of free neutrons; indeed, it is today their major source in burning shells of massive stars, although that insight is usually attributed to later emphasis by others.

It is unfortunate that Hoyle did not put to paper the equation he envisioned and described. Had he done so, clearer scientific visibility of his unparalleled achievement would have followed more easily.  Research in nucleosynthesis has followed his approach during five decades of countless numerical models evaluating Hoyle's equation. Nonetheless, the astrophysical world began to inappropriately cite a paper to appear, with Hoyle as coauthor, in 1957 for the general origin of nucleosynthesis in stars rather than Hoyle's works. Hoyle's great achievement lay somewhat obscurred until modern times, at least in terms of citations of research works.

\subsubsection*{Hans Suess, Nuclear Shell Model, and Abundances}
Hans Suess \index{Suess, H.} was born in Vienna in 1909 to a distinguished family of geologists. His chemical training and focus on abundances of the elements in meteorites enabled him to establish \emph{Suess's rules} for the abundances and their isotopes. The elemental abundances were not well known after WWII, but testing of a theory of nucleosynthesis needed that data base. In the late 1940s Suess began to argue that certain regularities of the abundances had to reflect nuclear properties of their isotopes. He used those systematics to coauthor with Hans Jensen a discovery paper \citep{Haxel49} for the magic numbers of nuclear shell structure. These are hugely important for nucleosynthesis. The magic numbers \index{nuclei!magic} are favored numbers of identical nucleons, either protons or neutrons, in the sense that clusters of those numbers of identical nucleons have larger than normal binding energy. They are 2, 8, 14, 20, 28, 50, 82 and 126, and they result from the combined effect of a deep spherical potential binding the nucleons together, coupled with a very strong spin-orbit energy that moderates energy gaps between differing nuclear  shells. For that paper Jensen shared the 1963 Nobel Prize in Physics with Maria Mayer for their independent  theoretical work on how the spin-orbit force establishes the magic numbers. Suess, working in Hamburg and Heidelberg, had seen the evidence of magic numbers in the abundance regularities shown by nuclei having a magic structure of neutrons or protons. For example, the abundances of the isotopes of barium implicated neutron irradiation in their nucleosynthesis. Suess amazingly divined that the very high abundance of $^{56}$Fe \index{isotopes!56Ni} had to reflect the doubly-magic properties of its $^{56}$Ni isobar having $N$=28 for both protons and neutrons. That correct assertion was not accepted for a decade because Hoyle's papers (and also B$^2$FH) \index{Burbidge, G.}\index{Fowler, W.A.}\index{Hoyle, F.}  had maintained that that abundance peak was established within a nuclear equilibrium having excess of neutrons relative to protons, so that the abundance of Fe isotopes could reflect their own nuclear properties rather than those of Ni. This error persisted for a decade, and had many dead-end astrophysical consequences. These have been intensively reviewed \citep{DDC99}, who also reviews Suess's pioneering papers from the 1940s.

The magic numbers were also to play a pivotal role in the theories for nucleosynthesis of the elements heavier than the Fe peak. Some of the earliest testaments to the correctness of the theoretical ideas lay in their interplay with the magic numbers. Suess himself called for two neutron-addition processes needed to account for magic-number abundances within the heavy elements. After moving to the United States in 1950, Suess coauthored with Harold Urey an immensely influential tabulation \citep{Suess56} of the abundances of the elements, largely from chemical analyses of meteorites rather than from astronomical observations. That review paper by Suess \& Urey \index{Urey, H.} became a cornerstone of empirical evidence for nucleosynthesis in stars.

\subsection{The Second Decade}
\label{sec:2-1-2}
	Hoyle's two papers were not enough to establish the theory of nucleosynthesis in stars. Contributions from other pioneers came together to launch the full theory and to win the acceptance that it enjoys today.

\subsubsection*{William A. Fowler and Burbidge, Burbidge, Fowler and Hoyle (B$^2$FH)}
William A. Fowler had met Fred Hoyle in 1953 when \index{Burbidge}\index{Fowler, W.A.}\index{Hoyle, F.}   Hoyle famously predicted the existence and the energy of the 0$^+$ second excited state of $^{12}$C. In that first application to nuclear physics of what has come to be called \index{anthropic principle} \emph{the anthropic principle}, Hoyle had argued that if such a nuclear state of $^{12}$C did not exist, neither would we! Fowler said that that prediction was what "really hooked me on nucleosynthesis".  Fowler met Hoyle again during his 1955 sabbatical leave in Cambridge UK. Fowler was at this time already the leader of the world effort (which Caltech championed) to determine the rates at which nuclear reactions would occur in stars. This involved the now familiar technique of measuring nuclear interactions at MeV energies with Van de Graff accelerator beams of charged ions, and extrapolating measured data downward in energy to the Gamow-peak energy within a thermal distribution \citep[][Ch.~4]{Clay68}. Fowler's zest for nuclear astrophysics was boundless, and he soon had the entire Kellogg Radiation Laboratory at work on thermonuclear reactions rates between positive ions. Fowler was awarded a share of the 1983 Nobel Prize in Physics for that pioneering program of research. His efforts had been directed primarily to the question of thermonuclear power in stars; but Fowler became intensely attracted also to the larger question of the origin of the elements. In Cambridge he met Geoffrey and Margaret Burbidge, and these three began a project with Fred Hoyle to write a survey paper on the issues of nucleosynthesis in stars. That paper was written at Caltech in 1956 and published the next year \citep{B2FH}. Soon called simply B$^2$FH, it became one of the most celebrated papers in astrophysics.

The authors of B$^2$FH contributed creatively and energetically to formulating the \emph{neutron-capture processes} \index{neutron capture} for synthesizing the elements heavier than nickel. Ascribing crucial roles to the magic neutron numbers $N$ = 50, 82 and 126 they described environmental situations in stars within which selected isotopes would be abundant, fleshing out ideas that Suess had envisioned earlier for two neutron-capture processes. Suess had, however, been unable to formulate these clearly enough to win contemporary acclaim. The slow neutron-capture process, named  \index{process!s process}  \emph{s~process} by B$^2$FH, envisioned 100-to-1000 years between neutron captures, so that radioactive isotopes would generally beta decay before capturing another neutron, keeping the capture path trailing along the valley of beta stability. The rapid-neutron-capture process, named \index{process!r process} \emph{r~process} by B$^2$FH, envisioned neutron densities so large that neutron captures occur in tens of milliseconds, faster than beta decays, with the result that the capture path moved into the realm of radioactive neutron-rich nuclei, being halted only when no additional neutrons could be stably added owing to their diminishing separation energies. These created \emph{waiting points}  \index{waiting points} at which the capture flow would halt and wait until beta decay occurs \citep[see also][]{Seeger65,Clay68}. Defining the $s$~process and the $r$~process was the high point of the B$^{2}$FH contributions to nucleosynthesis theory. Their Appendix included an inspiring table of all heavy isotopes in which each was characterized as being either $s$-process only, mostly s process, comparable $s$~process and $r$~process, mostly $r$~process and $r$-process only. This can be regarded as the next important step in nucleosynthesis theory.

An important aspect of astronomy with radioactivity lies in the competition between beta decay and neutron capture that ensues when neutron capture by stable isotopes create isotopes that can undergo beta decay. Most do so quickly in comparison with the time required to capture another neutron, but some key branch points are slow to decay. B$^2$FH \index{process!B2FH} had inventively shown that such competition at branch points could, when compared with the actual solar abundances, reveal the time scale and neutron density for the operation of the $s$~process. \index{branching point} Thermally populated excited states of radioactive nuclei often increase the effective decay rate, a delicate point in that aspect of astronomy with radioactivity.

Two cautions about B$^2$FH must be made in order to not overly eulogize what they achieved. Firstly, both the $s$~process and the $r$~process are, as described by
them, secondary processes of nucleosynthesis. \emph{Secondary nucleosynthesis} refers \index{isotopes!secondary} to the synthesis of new heavy nuclei from other existing heavy nuclei. It does not increase the galactic metallicity, a goal that Hoyle's founding papers had achieved spectacularly as the primary goal of the astronomy of nucleosynthesis. The B$^2$FH neutron-capture processes instead change one existing heavy nucleus into another. So those processes did not contribute to increasing metal abundances in the galaxy. Today it is known that, contrary to their description, the $r$~process is actually primary, because the collapsed supernova core synthesizes the seed nuclei that rapidly capture its free neutrons. The details of this are still not understood, however. Astronomical observations of old metal-poor stars confirm that the $r$~process indeed began earlier than did the $s$~process. Secondly, the B$^2$FH descriptions of the $s$~process and the $r$~process could not immediately be used for astrophysics calculations because they were time-independent formulations. Both the $s$~process and the $r$~process were described by the static condition $dN/dt=0$. The B$^2$FH descriptions were thus of stationary abundances that could exist within appropriate environments for these processes. This enabled a rough but clear correlation between the nuclear systematics of their process abundances and the solar isotopic abundances. But it did not enable calculation of the temporal growth of these abundances.

Over the next two decades B$^2$FH nonetheless became the default citation for workers wanting a reference to the general theory of nucleosynthesis in stars, vastly eclipsing the rate of citations to Hoyle's previous papers. One key to its success was its citation of over 100 stars showing nucleosynthesis effects in their spectra and of more than 100 astronomical research papers concerning those stars. This drew the astronomical community into the scientific culture of nucleosynthesis (which was then new to astronomers) as Hoyle's papers could not. Astronomers generally cited B$^2$FH rather than Hoyle, with the result that Hoyle's papers slipped into relative obscurity.

Fowler himself coauthored several other important works for nucleosynthesis, primarily with Fred Hoyle \index{Fowler, W.A.}\index{Hoyle, F.}  on supernovae and with Donald Clayton \index{Clayton, D.D.} on nucleosynthesis by neutron capture chains; however, his most important and essential role in history was the empirical thermonuclear reaction rates determined in his laboratory. For this he shared the 1983 Nobel Prize in physics. Probably Fowler's most significant subsequent work with Hoyle concerned the puzzling nature of the supernova phenomenon and on nucleosynthesis within them \citep{Hoyle60}. They divided supernovae \index{supernova!type I}\index{supernova!type II} into Types I and II based on theory, rather than on the presence of H absorption lines in the spectra. Type I was assigned to low-mass progenitor stars that evolved to white-dwarf stars of degenerate carbon, but then ultimately explode in an exothermic thermonuclear display that results in most of the iron in the universe. Electron-degenerate matter, supported by degeneracy pressure, is violently unstable to thermonuclear runaway. The Type II supernovae, on the other hand, occurs in massive stars whose cores are too massive to form white dwarf structure. In these continuing nuclear burning eventually exhausts the nuclear energy supply, and so the core must collapse. Fred Hoyle had predicted this inevitable collapse in 1954; but their 1960 paper carried the physical picture further. Today the world of astronomy uses this Hoyle-Fowler classification based on the physics of the device. It was a landmark in astrophysics \citep[see][]{Woosley99}.

\subsubsection*{A. G. W. Cameron, a Parallel Force}
The year 1957 also saw the emergence of A. G. W. Cameron \index{Cameron, A.G.W.} as one of the pioneers of nucleosynthesis. The Atomic Energy of Canada at Chalk River published in bound-mimeograph form a series of lectures on nucleosynthesis that Cameron delivered at Purdue University in March/April 1957 \citep{Cam57}. These lectures covered in an independent way essentially the same material as in B$^2$FH. Cameron had constructed his treatments by working alone, first as a new faculty member at Iowa State University and then at Chalk River, with only Hoyle's two papers to guide him as well as his training in nuclear physics from University of Saskatchewan. Owing to his independent treatment, Cameron's Lecture Notes became a valuable source of new ideas in nucleosynthesis. In his emphasis on nucleosynthesis within the separate shells in massive stars, Cameron's approach followed Hoyle's equation and thereby enlarged the ideas in \citet{Hoyle54} more effectively than did B$^2$FH. For the next two decades Cameron stressed in many publications with research students -- ones that he had recruited during a guest lecture course at Yale University -- the vast nucleosynthesis changes that occur in a massive star at its time of explosion. His papers establish him as the first great disciple of Hoyle's equation. Cameron's later Yale Lecture Notes (1963), written by his Yale students W.D. Arnett, C.J. Hansen and J.W. Truran, \index{Arnett, D.}\index{Truran, J.}\index{Hansen, C.J.} were much improved over \citet{Cam57}. They probably should have been published, but were not and are therefore not generally available; however, they were a strong influence on history, especially on Cameron's students and on Clayton and his students. Partly owing to their unavailability, Clayton published his own textbook \citep{Clay68}.

Cameron began his nucleosynthesis research, however, with the sources of free neutrons in stars that could be responsible for the observed radioactive Technetium \index{isotopes!98Tc}\index{Merrill, P.} observed by Paul Merrill in stellar atmospheres. Because all isotopes of the element Tc are radioactive, the presence of its absorption lines in stellar spectra argued that it had been created within the star during roughly its last lifetime. In 1955 Cameron proposed that the $^{13}$C($\alpha,n)^{16}$O reaction would liberate the extra bound neutron in the $^{13}$C nucleus and that the liberated neutron could be captured to create heavier isotopes \citep{Cam55}. Later  \citet{Cam59} calculated with the same motivation the number of neutrons liberated during carbon thermonuclear reactions following the exhaustion of He. The carbon fusion reactions were part of the sequence of thermonuclear stages within evolving massive stars (as \citet{Hoyle54} had first described).

In his large subsequent body of work, Cameron established himself as a true polymath. He introduced the speeding up of beta decay rate by thermal population of the excited states of a nucleus, many of which decay more rapidly than the ground state. During 1955-56 Cameron introduced numerical computation on the first vacuum-tube computers into nucleosynthesis problems, and remained thereafter on the cutting edge of nucleosynthesis computation. Cameron and his students repeatedly blazed new paths by programming nuclear reaction networks onto the latest and newest computers. He became an expert on planetary sciences and an important advisor to NASA. In particular, Cameron is forever famous for his work on the origin of the moon as the result of a Mars-like planetary collision with the young earth, a theory of origin that now seems beyond doubt.

\subsubsection*{Donald D. Clayton and Time-Dependent Heavy Element Nucleosynthesis}
Beginning as research student at Caltech with Fowler, Donald Clayton \index{Clayton, D.D.} began constructing a time-dependent formulation of the \index{process!s process} $s$~process in 1957, the year of B$^2$FH publication and of Cameron's Chalk River lecture notes. Clayton's discard of the assumption $dN/dt = 0$, an assumption requiring a constant $\sigma N$ curve for $s$-process abundances, altered profoundly the direction of $s$-process research by focusing on how efficiently seed nuclei could be converted to heavy $s$-process nuclei. B$^2$FH had not addressed that question. Clayton showed, as B$^2$FH had surmised, that the iron abundance peak must provide the seed nuclei being transmuted into the large overabundances of barium in stars whose spectra showed Ba/Fe some 20-50 times the solar ratio. More surprisingly his results also showed that, as cumulative neutron fluence increases, none of the sequential abundance distributions that are generated resemble the solar abundances \citep{Clay61A}.  The solar $s$-process abundances were required to be a superposition of differing numbers of Fe-seed nuclei (per Fe nucleus) exposed to differing integrated fluxes of free neutrons. The number exposed must be increasingly smaller for increasingly larger neutron irradiations. Galactic history or history within $s$-process stars is required to bring that superposition about. Therein lay new astrophysics. The solar $s$-process abundances were shown to not be simply a smoothly declining $\sigma N$ curve, as B$^2$FH had speculated, but a superposition of exposures generating narrow regions of atomic weight near the neutron-magic numbers where the assumption $dN/dt=0$ is severely violated. Two decades of improved measurements and consequent fitting to solar abundances \citep{Kaep82} would be required before advances in stellar evolution would be able to describe the exponential-like fluence distribution that was required. It was a sophisticated interplay between He-shell pulses and cyclically ignited H burning at the base of the envelope of AGB stars  \index{stars!AGB}  (see Chapter 3).

From the time of these first solutions of the neutron-irradiation superpositions resulting in the $s$-process abundances, new phenomenological aspects of heavy element nucleosynthesis were possible. The theory-based fit yielded all $s$-process \index{process!s process} abundances with meaningful accuracy. These allowed \citet{Clay61B} to publish the first decomposition of heavy-element abundances into their $s$-process and $r$-process parts. B$^2$FH had suggested the dominant processes for each isotope in their spectacular appendix; but a quantitative decomposition became possible for the first time. Their initial effort has been redone at least a dozen times as new neutron-capture-cross-section data appeared, most notably first by  \citet{Seeger65} and later by a new measurement program in Karlsruhe \citep{Kaep82}. This $s$-$r$ decomposition applied to astronomical spectroscopy of old stars has routinely produced meaningful new knowledge. Observations of old metal-poor dwarf stars indicated that the $r$-process abundances began to grow earlier as star formation first began than did the $s$-process abundances \citep{Tru81,Burris00}. That result demonstrated that the $r$~process is a primary nucleosynthesis process, rather than secondary as B$^2$FH had stated. This requires the $r$-process to occur within core-collapse supernovae.  Quantitative $s$-$r$ decomposition also inspired unforeseen new techniques for radioactivity-based cosmochronology.

Impressed by the new astrophysics lurking in time-dependence, Clayton advocated time-dependent formulation of the $r$~process as well, again jointly with Fowler and with P. A. Seeger\index{Seeger, P.A.}, Fowler's research student in Kellogg Lab. For the $r$~process as for the $s$~process, B$^2$FH had described only a time-independent steady flow that showed the neutron-rich heavy isotopes did indeed have abundant progenitors in suitable time-independent settings; but they had not been able to address whether the entire $r$-process \index{process!r process} abundances can be synthesized at one set of conditions. Computers at that time (1963) were not capable of handling a full $r$-process network. \citet{Seeger65} showed that the full mass range cannot be produced together (unless new seed nuclei are injected during the process). B$^2$FH had creatively defined the key nuclear physics relationships of the $r$~process but were mute on its dynamics. The time-dependent formulation by  \citet{Seeger65} became a prototype for $r$~process astrophysics. It showed the $r$~process to also be a superposition of differing irradiation histories of seed nuclei. Even after four decades of subsequent computations, the nature of the $r$~process superposition remains a frontier puzzle. It is noteworthy for astronomy with radioactivity that the entire $r$~process reactions occur within the realm of radioactive nuclei. Only after rapid expansive cooling can that neutron-rich radioactive abundance distribution undergo a series of beta decays changing each isobar identity until resting at the most-neutron-rich stable isobar (isobar is an isotope having the same atomic weight). A major research goal of nuclear astrophysics today is better laboratory definition of the parameters defining accurately the properties of the neutron-rich radioactive nuclei.

Understanding the time dependence during silicon burning (silicon photoerosion) \index{process!Si burning} was the big scientific challenge of the mid-1960s. Clayton introduced nuclear quasiequilibrium as a physical concept \citep{Bod68A,Bod68B} to clarify how silicon transmutes to an iron abundance peak. Quasiequilibrium explained the only big gap in Hoyle's 1954 theory of primary nucleosynthesis, replacing the ill-formulated $\alpha$ process of B$^2$FH.  A temporal sequence of quasiequilibrium states facilitated the calculation of the set $\Delta m_{k}$ for k=28-62 to be inserted in Hoyle's equation. The sequence of quasiequilibria again involved relaxing the assumption $dN/dt=0$. The quasiequilibrium concept was powerful and new, and enriched many subsequent aspects of nucleosynthesis reaction networks (the $r$~process, explosive oxygen burning, the $\alpha$-rich freeze out, the origin of \index{process!$\alpha$-rich freeze out} \index{isotopes!48Ca} $^{48}$Ca, and others). Just as all nuclear reactions proceed at the same rate as their inverses in full nuclear equilibrium (NSE), during quasiequilibrium one refractory nucleus violates that equilibrium assumption by changing abundance only slowly, while all others maintain equilibrium with it. During silicon burning the $^{28}$Si nucleus is the slowly changing, refractory post to which the \index{process!quasiequilibrium} quasiequilibrium distribution is attached \citep{Bod68A,Bod68B}. The most abundant isotopes between A=44 and 62 are, in this quasiequilibrium sequence, created as radioactive progenitors rather than as stable isobars, with important consequences for emerging astronomies of radioactivity.

\subsubsection*{The Sequel}

The decade 1956-66 following Hoyle's pioneering two papers had witnessed profound  enlargements of the theory of nucleosynthesis in stars. The years following the publication of B$^2$FH had been marked by vast improvement and reformulation of its influential processes. Calculable time-dependent descriptions of heavt-element nucleosynthesis processes refocused attention from simple correlations between nuclear properties and abundances to the astrophysical histories and stellar evolution that bring them into existence. The sometimes heard statement by astronomers that not much happened after B$^2$FH reflects lack of awareness of these historic changes. It may be of interest to note that the beginnings of nucleosynthesis theory was an international innovation. Of the pioneers named, only Fowler and Clayton were born Americans; Hoyle, E.M. Burbidge and G. R. Burbidge were English; Suess was Austrian; and Cameron was Canadian.

Innovations continued and accelerated during the next decade 1967-77. These will not be reviewed here except to say that the evaluation of Hoyle's equation through numerical computation of the evolution of massive stars yielded repeated insights into the interplay between stellar evolution and nucleosynthesis. The B$^2$FH neutron-capture processes revealed layers of complexity associated with the stars and with time dependences. The innovative center of this research moved away from Caltech, initially to Yale University and to Rice University, where Cameron and Clayton respectively founded schools evaluating Hoyle's equation. New leaders developed within those schools included especially W. David Arnett from Yale (subsequently also Rice) and Stanford E. Woosley from Rice (subsequently U.C. Santa Cruz). Each has many important publications on the evolution of massive stars and the nucleosynthesis in its shells \citep{Arn96,Woosley95} (see Chapter~4). A European role in this research also experienced rebirth during that decade, especially in Munich. It is also the decade 1967-77 that sees the emergence of several new observational aspects of astronomy with radioactivity. A description of those developments follows.

\subsection{New Astronomy with Radioactivity}
\label{sec:2-1-3}
	The existence of natural radioactivity clearly holds implications for the origins of atomic nuclei. The chemical elements could not have always existed if the radioactive nuclei were created along with the stable nuclei. In this way naturally occurring  radioactivity is intimately related to nucleosynthesis of the chemical elements. It was evident from the structure of the suggested processes of nucleosynthesis that radioactive nuclei played a large role in each of them. The key role of radioactive isotopes during nucleosynthesis and during thermonuclear power in stars was the first astronomy with radioactivity. Presumably the radioactive nuclei would be ejected from stars along with the new stable nuclei unless they decayed within the stars prior to ejection. Radioactivity plays a large role in the H-burning reactions, the PP chains and the CNO cycle \citep[see][]{Clay68} \index{process!pp}\index{process!CNO}  responsible for the stellar power capable of keeping the stars from cooling. Hans Bethe \index{Bethe, H.} was the 1968 Nobel Prize awardee for discovering these H-burning cycles in stars just prior to WWII. So it was evident in mid-20$^{\rm{th}}$ centrury that radioactive nuclei carry significant issues for astronomy. In the 1950s the idea arose of viewing the radioactivity at the solar center by detecting neutrinos \index{neutrino} arriving from the sun. Raymond Davis Jr. \index{Davis, R. Jr} would win a Nobel Prize for spearheading that effort.

	What made astronomy with radioactivity \index{astronomy} \index{radioactivity} so exciting scientifically was the discovery in the 1960s and 1970s of altogether new ways of observing radioactivity in astronomy. These aspects of astronomy with radioactivity generated new interdisciplinary connections to nucleosynthesis. All science needs observations to provide an empirical base, astronomy just as surely as laboratory science. To be sure, the scientific method happens differently in astronomy than in laboratory science. In astronomy there are no experiments that can change the initial conditions as a test of theories. No experiments can be planned to refute a hypothesis. One has instead only observations of natural events. In many cases simple observations can be sufficient to refute a hypothesis. It is fortunate that nature provides so many natural events, so that in many cases contrasting separate events constitutes a type of experimentation, giving a spectrum of observations in which conditions differ in ways that must be inferred but are nonetheless real. The experiments are natural and performed by nature herself, however, rather than by scientists.

	New types of observations of the occurrence of natural radioactivity galvanized astronomy by providing new kinds of astronomical data. These first historical advances are reviewed in what follows, as previews to the more detailed chapters that discuss them. They are: cosmochronology, the age of the elements; gamma-ray-line astronomy of radioactivity; radioactive power for light curves of supernovae and of novae; extinct radioactivity measured by excess abundances of daughter nuclei in solids formed in the early solar system; extinct radioactivity measured by excess abundances of daughter nuclei in solids condensed while dust particles condense as gas leaves a site of stellar nucleosynthesis. What follows is a brief description of how these applications were discovered or anticipated.


\subsection{Interpreting Exponential Decay}
The familiar exponential law of radioactive decay follows from Eq. (1.1) for an ensemble of $N$ radioactive nuclei from the constancy of decay probablility for a single nucleus. Given $N$ such nuclei the expected number of decays per unit time in the ensemble is given by the product of the number $N$ of nuclei and the decay probablility per unit time for a single nucleus. Accordingly

\begin{equation}\label{eq2}
\frac{dN}{dt} = N\lambda =N/\tau
\end{equation}

\noindent Because $\lambda$, and therefore $\tau$, is a constant for that nuclear species when it exists outside of stars, this equation has a well-known integral form,

\begin{equation}\label{eq3}
N(t) = N_0 \mbox{ } e^{-t/\tau}
\end{equation}

\noindent where $N_0$ is the initial number in the ensemble (at $t$=0). Similarly, if a number $N_0$ is observed now at $t=t_0$, the number that existed at an earlier time $t_1$ would have been \index{decay!radioactive} 

\begin{equation}\label{eq4}
N(t_1) = N_0 \mbox{ } e^{(t_0- t_1)/\tau}
\end{equation}

\noindent provided that new nuclei have not been added to the ensemble during that time interval. In the same spirit, if $N_1$ were created at earlier time $t_1$ and $N_2$ were added at a subsequent  time $t_2$, the number $N_0$ that would exist today at $t_0$ is \index{decay! half life}

\begin{equation}\label{eq5}
N_0 = N_1 \mbox{ } e^{-(t_0- t_1)/\tau} + N_2 \mbox{ } e^{-(t_0- t_2)/\tau}.
\end{equation}

Generalizing slightly, let $B_{\odot}(t)$ represent the fractional age distribution of the primary stable solar nuclei at the time of solar formation at $t = t_{\odot}$. Then $dB_{\odot}(t)/dt$ is the number of stable solar nuclei that had been born per unit time at time $t$. It is the age distribution that a radioactive species within an ensemble would have $if$ $it$ $were$ $not$ $decaying$. The age distribution for radioactive parents may be thought of as the age distribution of stable nuclei that were synthesized along with the radioactive nuclei\index{nucleocosmochronology}.  A mnemonic for the symbol $B_{\odot}$ is the $birthrate$ of those \index{stars!Sun} solar system nuclei. $dB_{\odot}/dt$ is the rate at which the stable nuclei and the primary parent radioactive nuclei were added to the total sample destined for the solar system rather than the rate at which they were produced by nucleosynthesis. In drawing this subtle distinction, \citet{Clay88} demonstrated the error of the common practice of equating the age distribution of solar nuclei with the rate of galactic nucleosynthesis. The age distribution of solar-system nuclei is vastly different from the galactic nucleosynthesis rate because the evolution of the ISM is involved.  Even if the ISM is instantaneously mixed, many of the old nuclei became trapped within stars, so that the age distribution in the ISM becomes biased toward more recent nucleosynthesis.

Given this definition of $dB_{\odot}/dt$ as the age distribution of solar-system nuclei, and given that their addition to the solar sample began at time $t_1$ and continued until $t_{\odot}$, the total number of any stable primary nucleosynthesis product would have been,

\begin{equation}\label{eq6}
N_{total} = \int_{t_1}^{t_{\odot}} \frac{dB_{\odot}(t')}{dt'} \mbox{ } dt'
\end{equation}

\noindent Then in analogy with Eq. \ref{eq5}, if those nuclei are instead radioactive, the number surviving until solar formation at time $t_{\odot}$ would be

\begin{equation}\label{eq7}
N_{\odot} = \int_{t_1}^{t_{\odot}} \frac{dB_{\odot}}{dt'} \mbox{ } e^{-(t_{\odot}-t' )/\tau} dt'
\end{equation}

\noindent If the total duration $t_0 - t_1$ of production is much greater than the mean lifetime $\tau$, and if $dB_{\odot}$/dt varies only slowly, Eq. \ref{eq7} reduces approximately to

\begin{equation}\label{eq8}
N_{\odot} = \left( \frac{dB_{\odot}}{dt}\right)_{t_{\odot}} \tau
\end{equation}

\noindent This useful formula estimates the number of remaining radioactive nuclei to be equal to  the number created during the last mean lifetime of that radioactive species. If $dB_{\odot}/dt$ has jagged temporal structure near $t_{\odot}$, however, a more complicated evaluation would be required. The Eq. \ref{eq8} is very useful as a first approximation to the numbers of radioactive nuclei within the ISM during continuous nucleosynthesis in stars if the birthrate $B(t)$ is redefined as the birthrate spectrum of those stable ISM nuclei residing in the ISM rather than in stars.

	These relationships expressing properties of exponential decay are central to understanding both extinct radioactivity in the early solar system and the numbers of radioactive nuclei within astronomical objects.

\section{Disciplines of Astronomy with Radioactivity}
\label{sec:2-2}

\subsection{Nuclear Cosmochronology}
\label{sec:2-2-1}

Today it is self evident that the existence of radioactive nuclei implies that they were created at some estimable moment in the past. Were that not the case, they would have long ago have disappeared. But the full implications were not evident to those engaged in the exciting day-to-day goals of understanding natural radioactivity. Nonetheless, in 1929 Ernest Rutherford \index{Rutherford, E.}  \index{nucleocosmochronology} wrote what may be the first paper on astronomy with radioactivity \citep{Ruth29}. Today we use the term nuclear cosmochronology to mean the attempt to use natural abundances of radioactive nuclei to compute the age of an astronomical object or of the elements themselves. Determining the age of the chemical elements by assuming them to have the same age as the radioactive nuclei became a goal that has attracted many, most notably William A. Fowler and this writer.

\subsubsection*{Uranium and Thorium on Earth}
When he began to think on these things, Ernest Rutherford concluded from the ratios of $^{235}$U/$^{238}$U \index{isotopes!235U}\index{isotopes!238U} as measured in his laboratory that uranium was created somehow within the sun and transported to earth. Accepting Jeans's estimate that the age of the sun was an exuberant 7$\times$10$^{12}$ years, he wrote in a prophetic paper addressing the puzzle \citep{Ruth29}:

\begin{quote}
{\it ..it is clear that the uranium isotopes which we observe on earth must have been forming in the sun at a late period of its history, namely, about 4$\times$10$^9$ years ago. If the uranium could only be formed under special conditions in the early history of the sun, the actino-uranium on account of its average shorter life would have practically disappeared long ago. We may thus conclude, I think with some confidence, that the processes of production of elements like uranium were certainly taking place in the sun 4$\times$10$^9$ years ago and probably still continue today.}
\end{quote}

Corrected modern numbers for those that Rutherford used are the observed
abundance ratio $^{235}$U/$^{238}$U=7.25$\times$10$^{-3}$ and the mean lifetimes against alpha decay, $\tau$($^{235}$U)= 1.029$\times$10$^9$ yr and $\tau$($^{238}$U)= 6.506$\times$10$^9$ yr. Since each isotopic abundance has been exponentially decaying during the age $A_E$ of the earth, their abundance ratio when earth formed would have been ($^{235}$U/$^{238}$U)$_0$ = 7.25$\times$10$^{-3}$ ($e^{A_E/\tau(^{235}\rm{U})}/e^{A_E/\tau(^{235}\rm{U})}$).

The age of the earth has been reliably measured using the fact that these two isotopes of U come to rest, after a series of alpha decays, as different isotopes of Pb, namely $^{207}$Pb and $^{206}$Pb respectively. Using the measured earth age, $A_E$= 4.57$\times$10$^9$ yr, yields the initial U ratio on earth to have been ($^{235}$U/$^{238}$U)$_O$ = 0.31. These facts are beyond doubt.

	The implication of great consequence stems from the expectation that the $r$~process that is responsible for the nucleosynthesis of both isotopes, should make more
$^{235}$U than $^{238}$U. $^{235}$U has six non-fissioning progenitors whereas $^{238}$U has but three. An $r$-process progenitor is a non-fissioning transuranic nucleus that after a series of relatively fast radioactive decays comes to rest at one of these long-lived isotopes of U. For $^{235}$U those nuclei are $A$=235, 239, 243, 247, 251 and 255-totaling six $r$-process progenitors-whereas for $^{238}$U they are $A$= 238, 242, 246 and 35\% of 250-totaling 3.35 progenitors. Taking into account the empirical evidence that production by the $r$-process favors even-$A$ nuclei by a slight 20\% over adjacent odd-$A$ nuclei, one expects $^{235}$U production to exceed that of $^{238}$U by a factor near $P(^{235}$U)/$P(^{238}$U)=1.79. This argument reveals that the abundance ratio $^{235}$U/$^{238}$U has declined from near 1.79 at production to 0.31 when the earth formed. This decline takes considerable time, showing that U isotopes were synthesized during pre-earth astrophysical history.

	This first calculation of nuclear cosmochronology reveals the nature of the problem, but also its uncertainties. Is the production ratio $P(^{235}$U)/$P(^{238}$U)=1.79 correctly estimated? Were the U isotopes synthesized in one single presolar event, in which case it occurred about 6.6 billion years ago, 2 billion years prior to formation of the earth. Or was their production rate distributed in presolar time? If the age distribution of the solar system's $r$-process nuclei is flat between the time of first production and the earth's formation, the beginning of nucleosynthesis would fall near 13 billion years ago. The true age distribution can hardly be known with any assurance, revealing the severe limitation of this single pair for constraining the time of the beginning of nucleosynthesis in our galaxy.

The technique was extended to the ratio $^{232}$Th/$^{238}$U \index{isotopes!232Th} by  \citet{Fow60}; and Fowler returned to it many times in later years  \citep[e.g.][]{Fow72}; see also the textbook by \citet{Clay68}. Relevant numbers used are the observed abundance ratio $^{232}$Th/$^{238}$U= 4.0 and the mean lifetimes against alpha decay, $\tau$($^{232}$Th)= 20.04$\times$10$^9$ yr and $\tau$($^{238}$U)= 6.506$\times$10$^9$ yr. The abundance ratio is much more uncertain than it was for the pair $^{235}$U/$^{238}$U because Th and U are different chemical elements. Any two elements having different fractionation chemistry pose a tough problem when seeking their ratio in the initial solar system (or in the sun).  Since each isotopic abundance has been exponentially decaying during the age $A_E$ of the earth, their abundance ratio when earth formed would have been ($^{232}$Th/$^{238}$U)0=2.5 instead of 4.0. And the production ratio in $r$-process events was inferred from the same counting arguments of odd and even progenitors of $^{232}$Th and $^{238}$U to be $P(^{232}$Th)/$P(^{238}$U) = 1.73.

	A problem is that this Th/U pair does not give transparently concordant numbers with the U pair. Uncertainties in $r$-process production ratios, in the relative abundances of Th and U, and in the arbitrary parameterizations of galactic chemical evolution that have been used each conspire to yield possible solutions in which nucleosynthesis began anywhere from 2 to 10 Gyr prior to solar birth.  Fowler strove repeatedly to circumvent these uncertainties, trying to extract the true answer (for there obviously is a true answer!); but in fact, the data are not adequate and the astrophysical model of the galaxy used is not adequate for the task. Data from other chronological species would be needed, along with a more sophisticated appreciation of galactic chemical abundance evolution.

	On a more positive note, Fowler's papers \citep[][and several others]{Fow60,Fow60B,Fow72}  inspired many others to tackle this fascinating aspect of astronomy with radioactivity.

\subsubsection*{Cosmochronology with Extinct Radioactive Abundances}
	In 1960s a new discovery, excess trapped $^{129}$Xe \index{isotopes!129Xe} \index{meteorites} gas in meteorites \citep{Rey60}, provided the first evidence that the matter from which the solar system formed contained radioactive nuclei whose halflives are too short to be able to survive from that time until today. These are called extinct radioactivity.
	Discovery of extinct radioactivity utilized the buildup of the daughter abundance from a radioactive decay as a measure of how abundant that radioactive parent was in the \index{solar system} initial solar system. What  \citet{Rey60} showed was that the excess $^{129}$Xe gas in meteorites had resulted from $^{129}$I decay and that the initial $^{129}$I abundance was about 10$^{-4}$ of the initial $^{127}$I \index{isotopes!127I} \index{isotopes!129I} abundance. He realized that this datum allowed an estimate of when the iodine isotopes had been created. By assuming that at the time of their nucleosynthesis the initial $^{129}$I abundance had been equal to that of stable initial $^{127}$I, Reynolds argued from the mean lifetime $\tau$($^{129}$I)= 23.5$\times$10$^6$ yr (23.5 My) that iodine had been created only about 300~My prior to solar system birth. This was the first new cosmochronology technique since Rutherford and brought Reynolds great fame. Unfortunately, Reynolds' conclusion was no more believable than Rutherford's had been, because his astrophysical model was also quite unrealistic. He mistakenly assumed that all of the iodine abundance in the solar system had been created at the same time, and therefore used Eq. \ref{eq3} to estimate that time. Because 10$^{-4}$ requires a decay period equal to about 13 mean lifetimes, he had concluded that the age of the elements was 13$\tau$($^{129}$I)= 300~My.   \citet{Fow60B} countered that it was more realistic to assume that the synthesis of iodine isotopes was spread out by multiple $r$-process events, occurring between the time t$_1$ of a first nucleosynthesis event until the time $t_0$ of the last such event prior to solar birth. That concept, called $continuous$ $nucleosynthesis$, required use of the Eq. \ref{eq8} limit of Eq. \ref{eq7} instead of Eq. \ref{eq3}. That calculation gave the initial nucleosynthesis epoch to be 104$\tau$($^{129}$I)~=~230~Gy -- very old indeed, unrealistically old. This picture was then modified to allow for an interval of sequestered interstellar gas, probably within a molecular cloud, of about 100~My during which no new nuclei were added to the solar mix -- a so-called $free$ $decay$ $interval$ in which radioactive abundances would follow Eq. \ref{eq3}. Such an interval would require that 100~My before the solar birth, when that interstellar gas withdrew from new nucleosynthesis, the interstellar abundance ratio would have been $N(^{129}$I)/$N(^{127}$I) = 2.8$\times$10$^{-3}$ rather than 10$^{-4}$.  Eq. \ref{eq8} then suggests that nucleosynthesis began 8~Gy prior to solar birth, a reasonable number, obtained, unfortunately, by construction. But that construction had nonetheless established a new paradigm for radioactive abundances in the early solar system. The upshot for the problem of radioactive chronology is that the abundance of now extinct radioactivity in the early solar system can say little about when nucleosynthesis first began. But it provides other equally interesting issues for the astronomy of radioactivity.

	Radioactive $^{129}$I is of special historical significance as the first extinct radioactive nucleus to be discovered that can be regarded as part of the average radioactivity in the galactic ISM at the time of solar birth. If so it is in that sense typical of what is expected in the ISM. Other extinct radioactivities may not be capable of such an interpretation but must instead be interpreted as produced by a special event associated with the birth of the solar system. Other subsequently discovered extinct nuclei that can be regarded in the first category of galactic survivors are $^{53}$Mn, $^{107}$Pd, $^{182}$Hf, $^{146}$Sm and $^{244}$Pu.

\subsubsection*{Birthrate Function for Primary Solar Abundances}
	The history of nuclear cosmochronology reveals widespread confusion between the rate of galactic nucleosynthesis and the age spectrum of solar system nuclei. It is the latter that enters into cosmochronology through Eq. \ref{eq7}. So endemic is this confusion between these two fundamental concepts that it is important to define in this section the birthrate spectrum for primary abundances in the solar system.

Let $N_{\odot}$ be the solar abundance of a primary nucleosynthesis product.  One may think of it either as the total number of those atoms in the solar system, which consists of awkwardly large numbers, or as the total number of those atoms in the solar system normalized to a defined abundance for a specified nucleus that sets the abundance scale. Two such normalized scales are very common: one used more in astronomy defines the solar abundance of hydrogen to be $N_H$ = 10$^{12}$; another, used more in solar abundances derived from meteorites, defines the solar abundance of silicon as $N_{Si}$=10$^6$. This book will primarily use the latter definition because most studies of nucleosynthesis in the literature use the scale Si=10$^6$ (where the chemical symbol is often used to represent the abundance of that element).  Let $B_{\odot}(t)$ be the cumulative fraction, as a function of time, of stable solar primary nuclei that already existed at time $t$. Letting for convenience the starting time $t=0$ be the time when nucleosynthesis first began, the function $B_{\odot}(t)$ starts at 0 at $t$=0 and rises to $B_{\odot}$=1 at $t=t_0$, the time when the solar system formed. Note carefully, $B_{\odot}(t)$ is not the cumulative birth fraction of that nucleus in galactic nucleosynthesis, but rather the cumulative birth fraction of those primary nuclei that actually entered the solar system inventory. Normalizing this definition of $B_{\odot}(t)$ as a function that rises to unity at solar birth, the product

\begin{equation}\label{eq9}
N_{\odot} \cdot B_{\odot}(t) = N(t)
\end{equation}

\noindent where $N_{\odot}$ is the solar abundance of that stable species, $N(t)$ is the number of of those solar-system nuclei that already existed at time $t$. It is equally clear that $dB_{\odot}(t)/dt$ is the fractional birthrate per unit time of primary solar nuclei. It is the age distribution of solar nuclei, and redefined as here to rise to unity it is approximately  the same function for all primary stable nuclei.

	A mental experiment is needed to set this birthrate function \index{chemical evolution!cosmic} clearly in mind before considering why it differs so from the nucleosynthesis rate. The reader is warned that the research literature suffers endemic confusion over that distinction, requiring its careful definition. For any primary nucleus within solar abundances, \emph{paint each solar-system atom of it red!} This cannot of course be done in reality, but the thought experiment helps understanding of the birthrate spectrum. On the scale Si=10$^6$ one has by construction 10$^6$ \emph{red} Si atoms in the solar system. Then imagine watching a film run backwards in time of these 10$^6$ \emph{red} Si atoms. Back in the interstellar medium they are mixed with a much larger number of \emph{unpainted} Si atoms. As the film runs to earlier times, these 10$^6$ \emph{red} Si atoms have greatly differing histories. Some, after being created in a supernova, have later been inside a star, survived and come back out. But for each atom, its world line reaches a time when that atom first came into existence. Each of the 10$^6$ red solar Si atoms is labeled by that birthdate, born from distinct supernovae at differing times. Then a graph is constructed giving the total number that existed at time $t$. This graph starts at 1 atom when the first solar Si atom is created and reaches 10$^6$ by the time the solar system forms, when all 10$^6$ have been created. If the curve is then normalized by dividing by 10$^6$, one has the birthrate spectrum $B_{\odot}(t)$ of solar primary nuclei, which rises to unity at the time the solar system forms. Eq. \ref{eq9} depicts this situation, where $N(t)$ rises to 10$^6$ at solar birth, but $B_{\odot}(t)$ rises to unity at that time. Although different primary species have vastly differing abundances, for each the function $B_{\odot}(t)$ is approximately the same function because the abundances of all primary nucleosynthesis products rise approximately (but not exactly) together.

	Eq. \ref{eq9} is the solution of the differential equation
\begin{equation}
dN/dt = dB_{\odot}(t)/dt  {\rm \mbox {  }for\mbox {  }stable\mbox {  } nuclei}
\end{equation}
and
\begin{equation}\label{eq10b}
 dN/dt = dB_{\odot}(t)/dt - \lambda N \mbox{    }	\rm{for} \mbox{ } \rm{radioactive} \mbox{ } \rm{nuclei}
\end{equation}

\noindent
One frequently sees these equations written for galactic nucleosynthesis, with $dB/dt$ identified as the galactic nucleosynthesis rate and
$N(t)$ as the ISM abundance. This is a fundamental error. Consider why. \index{chemical evolution!galactic}

 The galactic nucleosynthesis rate for Si atoms was very large in the young galaxy, when the rate of formation of massive stars and supernovae was large. But those new Si atoms in the interstellar medium become mostly locked up in the interiors of subsequent low-mass stars, and are not available for incorporation into the solar system. This occurs because it is the fate of most interstellar gas to be locked up in low-mass stars. The initially gaseous galaxy is today about 10\% gas, with 90\% trapped in low-mass stars. The origin of the Si atoms that entered solar matter is therefore biased towards supernovae that occur relatively shortly before solar birth, and biased against those that were created long earlier. This important conceptual distinction was demonstrated quantitatively by \citet{Clay88}, who stressed the concept of the age spectrum of solar nuclei and who constructed analytic models of galactic chemical evolution that enable a full disclosure of such issues. As an example, the simplest of all realistic models supports a star formation rate and supernova rate that declines exponentially as the interstellar gas is exponentially consumed; but the age spectrum of solar nuclei in that model is constant -- equal numbers from equal times, despite the strong bias of nucleosynthesis of Si toward early galactic times. The student can study  \citet{Clay88} for many related issues for cosmochronology. Some of these issues involving galactic chemical evolution will be addressed later.

\subsubsection*{Uranium and Thorium in old Dwarf Stars}
	  An altogether new technique in radioactive chronology \index{nucleocosmochronology} became possible as CCD detectors enabled astronomers to measure line strengths for much weaker lines than had been previously possible. \citet{But87} advanced the first such argument when he was able to accurately measure the abundance of thorium in old dwarf stars. The measured abundances enabled him to argue that the oldest stars are no older than 10~Gy. If they were, Butcher argued, the Th, which has resided in the old stars since their births, would by now have decayed to a smaller abundance than it is observed to have. Because the dwarf stars observed appear to be among the early stars formed in our galaxy, the argument concludes that the galaxy age is negligibly greater than 10~Gy. It will be clear that since the $^{232}$Th \index{isotopes!232Th} mean life is a very long 20~Gy, it can have decayed from its initial abundance by only $e^{-0.5}=0.6$ during a 10~Gy life of a dwarf star. Therefore the method requires not only an accurate measurement of the Th abundance but also an accurate estimate of its initial abundance when the star formed. It is the reasoning required to obtain that initial abundance that is a controversial aspect of this method. Realize also that the Th in an old star is decaying freely, so that its abundance declines faster than in the ISM where new nucleosynthesis of Th continuously replenishes it.

	The method advocated measures the ratio of Th/Nd in the star and observes that the measured ratio is smaller than that seen in younger stars. It assumes that the initial Th/Nd ratio in the star would be the same as the ratio seen in young stars, since both Th and Nd are products of the $r$~process and may be hoped to have a constant production ratio there. It was then counter argued \citep{Clay87B,Clay88} that a constant production ratio for Th/Nd is not to be expected because almost half of the solar Nd abundance has been created in the \index{process!s process} $s$~process, and, furthermore, the $r$~process is primary whereas the $s$~process is secondary. There exists ample and exciting evidence from the observed  $r$-process pattern of abundances in extremely metal-deficient old stars \citep{Tru81,Gil88,Beers05} that the $r$~process \index{process!r process}nucleosynthesis began prior to the beginning of the $s$~process. That seemed to support the initial skepticism about Butcher's technique.  But astronomers hurried to point out that if one omits the very metal-deficient old stars, the ratio of $r$-to-$s$ abundances in stars having more than 10\% of solar abundances is observed to be a near constant.  \citet{Clay88} revisited that larger puzzle of parallel growth of $s$ and $r$ nuclei, noting that the abundance evidence suggests that the $s$~process is primary despite its building upon iron seed nuclei; moreover, he presented a nucleosynthesis argument that showed how the $s$~process could in fact resemble primary nucleosynthesis even though it is secondary! See section 3.1 of \citet{Clay88} for that argument, which has proven to be of high significance for nucleosynthesis theory.

	The lesson to be taken from this exciting new technique and its controversy is this: radioactive nuclear cosmochronology is vitally dependent upon a correct picture of the chemical evolution of the galaxy. As such it has become less the province of nuclear physics and more the province of astronomy. Tinsley and Clayton both made that point in timely and influential ways.  Only when many essential details of the history of our galaxy and of the history of nucleosynthesis within it have been settled can these techniques of radioactive chronology yield a reliable answer for the age of the chemical elements. Nuclear cosmochronology is truly an astronomy with radioactivity.

\subsubsection*{Cosmoradiogenic Chronologies}
	In the early 1960s a different approach to radioactive chronology became possible. It utilizes the buildup of the daughter abundances of radioactive decay during the history of interstellar matter as a measure of how long that decay had been occurring and, therefore, how long ago the production of radioactivity began. One can imagine the stable daughter of a radioactive decay as a bucket into which all interstellar decay of the radioactive parent has been collected. It integrates past decay rather than focusing on how much radioactivity remains. Such daughter-isotope buildup during the history of the earth was already known as {\it radiogenic abundance}. It had been applied to the ages of earth and of meteorites, samples in which it could be expected that the other initial isotopic compositions were well known-namely, the solar abundances. But radiogenic abundance collected during the history of interstellar matter, what  \citet{Clay64} called {\it cosmoradiogenic abundance}, was not seen as possible data for determining the age of the elements themselves because the interstellar abundances are increased by the processes of nucleosynthesis as well as by any radioactive decay for specific isotopes. Thus the cosmoradiogenic abundance could not be easily disentangled from the direct nucleosynthesis abundance.

	What cut through that impasse was a credible {\it quantitative theory} of heavy-element nucleosynthesis. Suess, B$^2$FH and Cameron had each contributed importantly to the idea that two distinct neutron-capture processes, the $s$~process and the $r$~process, had been  responsible for the creation of all but the very lightest isotopes of the elements heavier than about $Z$=32. The disentanglement required two things: firstly, an accurate $s$-process theory and secondly, a parent-radioactivity abundance that could be produced only by the $r$~process and a daughter that could be produced only by the $s$~process. The radiogenic daughter of such a radioactive $r$-process isotope is called a shielded isotope, because it is shielded from $r$-process production by the radioactive parent.

 \citet{Clay64} made the key first step by noticing that the solar abundance of $^{187}$Os \index{isotopes!187Os} \index{isotopes!187Re} \index{nucleocosmochronology} is about twice as great as it is expected to be from $s$-process nucleosynthesis, and that it cannot be synthesized by the $r$~process because neutron-rich matter at $A$=187 will, upon decaying toward stability, arrest at $^{187}$Re. The decay cannot reach $^{187}$Os, which is shielded by $^{187}$Re. The $r$~process production flows into $^{187}$Re. The abundance of $^{187}$Os can then be thought of as having two parts; a part produced by the $s$~process and a part produced by the decay of very-long-lived $^{187}$Re. The quantitative $s$-process theory \citep{Clay61A} reliably accounted for about half of the $^{187}$Os abundance, so the other half had to be the result of the beta decay of $^{187}$Re during ISM residence:

\begin{center}
$^{187}\rm{Re} \longrightarrow ^{187}\rm{Os} + e^- + \nu $ \mbox{       }	 $ t_{1/2}= 43\ \rm{Gyr}$
\end{center}

\noindent That the halflife exceeds the galaxy's age is useful, because only a modest fraction of the $^{187}$Re can therefore have decayed during presolar history and most of it therefore still exists; but because $^{187}$Re is six times more abundant than the $s$-process amount of $^{187}$Os, a decay of only 1/6$^{\rm{th}}$ of the $^{187}$Re can have doubled the $s$-process $^{187}$Os abundance. The persistence of live $^{187}$Re is even necessary for the method, because it allows one to equate the total amount of $^{187}$Re nucleosynthesis that was destined for solar incorporation to the sum of the quantity of $^{187}$Re remaining at solar birth and the quantity of $^{187}$Re that had already decayed prior to solar birth. The latter is the difference between the total solar abundance of daughter $^{187}$Os and the quantity produced by the $s$~process. That is,

\begin{center}
$^{187}\rm{Os} = ^{187}\rm{Os}_s + ^{187}\rm{Os}_c$
\end{center}

\noindent where $^{187}$Os$_c$ is the cosmoradiogenic part of the $^{187}$Os  abundance owing to cosmic $^{187}$Re decay. And the $s$-process part can be obtained from $^{187}$Os$_s$ = $^{186}$Os ($\sigma$(186)/$\sigma$(187)), where $\sigma$ is the appropriate neutron-capture cross section during the operation of the\index{process!s process} $s$~process. This last relationship is valid because the entirety of $^{186}$Os is from the $s$~process because it is shielded from $r$-process\index{process!r process} nucleosynthesis.

	The discovery of the Re-Os clock was the key that \citet{Clay64} used in presenting solutions to three cosmoradiogenic chronologies. These are the beta decay of $^{187}$Re to $^{187}$Os described above, the similar beta decay of $^{87}$Rb to $^{87}$Sr, and the $\alpha$-decay chains by which $^{235}$,$^{238}$U decay to $^{207}$,$^{206}$Pb. Each had been known in the study of meteorites, but by this work they joined studies of the age of the elements themselves.

	Cosmoradiogenic lead is the more interesting of the other two chronologies, in part because it couples cosmoradiogenic Pb to the older chronology based on $^{235}$U/$^{238}$U, and in part because the Pb isotopes \index{isotopes!206Pb} \index{isotopes!207Pb}  have three contributions to their abundances:

\begin{center}
\noindent	$^{206}$Pb = $^{206}$Pb$_s$ + $^{206}$Pb$_r$ + $^{206}$Pb$_c$

\noindent	$^{207}$Pb = $^{207}$Pb$_s$ + $^{207}$Pb$_r$ + $^{207}$Pb$_c$
\end{center}

\noindent where $^{206}$Pb$_c$  is the cosmoradiogenic part from $^{238}$U \index{isotopes!238U} decay, and where $^{207}$Pb$_c$  is the cosmoradiogenic part from $^{235}$U decay. Suffice it to be said here that the $s$-process part and the $r$-process part can both be estimated, and from them the chronological solutions can be displayed. Readers can turn to \citet{Clay88} for these solutions.

	The cosmoradiogenic chronologies seem to indicate an older galaxy than do the direct-remainder chronologies \citep{Clay88}. But they also are compromised by unique and interesting uncertainties. For the $^{187}$Re cosmoradiogenic chronology, the main uncertainty is how greatly its decay rate is speeded by its incorporation into stars, where ionization increases its beta decay rate. \citet{Yok83} evaluated this effect within an ambitious galactic chemical evolution model that enabled them to take into account the fraction of $^{187}$Re that is incorporated into stars and ejected again without nuclear processing and how much time such interstellar $^{187}$Re spends inside of stars. Their results can be reproduced by a 40\% increase of the neutral $^{187}$Re decay rate. For the Pb cosmoradiogenic chronology, the main uncertainty is an especially interesting set of nucleosynthesis problems associated with both the $s$-process part and the $r$-process part of their abundances. Uncertainty about the Pb/U elemental abundance ratio also suggests some caution.

	Certainly it can be said that these chronologies present intricate problems in the astronomy with radioactivity.

\subsubsection*{Radioactivity and Galactic Chemical Evolution}

In astronomy, the abundances of radioactive species are interpreted through the lens of \index{chemical evolution!galactic} galactic abundance evolution, frequently called galactic chemical evolution (GCE). Interpretations of nuclear cosmochronology or of initial solar abundances of extinct radioactivities depend on the mean expected abundances in the ISM when solar birth occurred.  Galactic Chemical Evolution traditionally concerns the chemical composition of the mean ISM. It features continuous nucleosynthesis between a starting time and the time when solar formation occurred. During that lengthy galactic period radioactive abundances undergo decay and are incorporated into new stars, but they are also replenished by the injection of freshly synthesized radioactivity and are diluted by low-metallicity matter falling onto the disk. Competition among these terms renders simple galactic chemical evolution a showplace for the behavior of radioactivity.
One differential equation describing that abundance of those specific galactic atoms (the \emph{red atoms}) that will later be incorporated into the solar system can be thought of as the solution of
\begin{equation}\label{eq10}
\frac{dN}{dt} = \frac{dB_{\odot}(t)}{dt} - \frac{N(t)}{\tau}
\end{equation}
\noindent
where $\tau$=1/$\lambda$ is the mean radioactive lifetime of the nucleus in question, N(t) is its time-dependent abundance within those (red) atoms that are destined for inclusion in the solar system, and $B_{\odot}(t)$ is the birthrate spectrum of those solar nuclei, defined such that $\frac{dB_{\odot}(t)}{dt}dt$ is the number of stable solar nuclei that were born between times t and t+dt. The birthrate spectrum of solar nuclei is unknown, however, so eq. \ref{eq10} can not be solved. Despite this limitation, workers persisted in attempting to fix the time of the beginning of nucleosynthesis from eq.\ref{eq7}, which students can show does solve the differential equation \ref{eq10} above. Many tried this approach by assuming an easily integrable form for $\frac{dB_{\odot}(t)}{dt}$, most often $exp(-Gt)$, where $G$ is taken (erroneously) to be an unknown positive number. That assumption proved seductive to many, because the rate of galactic nucleosynthesis is believed to decrease smoothly with time, so maybe $\frac{dB_{\odot}(t)}{dt}$ can be assumed to decline as well. Importantly, however,$\frac{dB_{\odot}(t)}{dt}$ differs greatly from the galactic production rate and actually grows with time. The reason for this surprise requires understanding.

The birthrate spectrum imagines that the nuclei destined for the solar system could be tagged at birth (imagine red paint) and ignores all other ISM nuclei. Because $B_{\odot}(t)$ is unknown, however, astrophysicists instead calculate the evolution of the interstellar abundances. Their mean values in a well mixed ISM at the time of solar birth provide the expected solar abundances. The conceptual difference between $B_{\odot}(t)$ and the galactic nucleosynthesis rate has confused many unwary researchers. Many published papers have used their chosen form for $\frac{dB_{\odot}(t)}{dt}$ and integrated eq.\ref{eq7} in the hope of fixing the beginning time of galactic nucleosynthesis (the lower limit) by comparing those calculated results with known initial radioactive abundances in the solar system. A common form because eq.\ref{eq7} is then easily integrated is $\frac{dB_{\odot}(t)}{dt}  = exp(-Gt)$. In retrospect, however, such poor choices for the form of $\frac{dB_{\odot}(t)}{dt}$ render their conclusions invalid.

How does GCE clarify the problem of galactic radioactivity? The mass $M_G$ of ISM gas is reduced owing to the rate $\psi(t)$ of its incorporation into new stars, but it is increased by the rate $E(t)$ at which mass is ejected from old stars and by the rate $f(t)$ at which new mass falls onto the galactic disk:
\begin{equation}\label{eq11}
\frac{dM_G}{dt} = - \psi(t) +E(t) + f(t)
\end{equation}
If one takes the ejecta rate $E(t)$ from the spectrum of newly born stars to be a fixed return fraction $R$ of the rate $\psi(t)$ at which mass joins new stars, and if one assumes linear models in which the rate of star formation is proportional to the mass of gas $M_G$ , the equations governing the interstellar composition of both stable and radioactive nuclei can be solved analytically within families of choices for the infall rate $f(t)$, as \citet{Clay88,Clay85} has shown. These analytic solutions for ISM abundances do indeed clarify radioactive abundances. For that purpose Clayton writes
\begin{equation}\label{eq12}
\frac{dM_G}{dt} = - \psi(t)\mbox{ }(1-R)\mbox{ }+ f(t) = -\omega M_G + f(t)
\end{equation}

\noindent where $\omega M_G(t)  = (1-R)\psi(t)$ is valid for linear models.  The constant $\omega$ is the rate of consumption of gas by star formation when compensated by gaseous return from stars. These are called \emph{linear models} because the star formation rate $\psi(t)$ is taken to be proportional to the mass $M_G(t)$ of interstellar gas. That linear assumption is not strictly true; but it is plausible in taking the star formation rate to increase as the mass of gas increases, and to decline as it declines. Furthermore, it is supported by observations of star formation rates in spiral galaxies. From their observations, \citet{Gao04}  state, ``The global star formation rate is linearly proportional to the mass of dense molecular gas in spiral galaxies.'' The purpose of models of galactic chemical evolution insofar as galactic radioactivity is concerned is to understand the mean expectation for radioactivity in the ISM. The families of linear analytic models are constructed for that purpose.

	Coupled with eq. \ref{eq12} is an equation for the rate of increase of the concentration $Z$ of each interstellar nucleosynthesis product. The concentration is defined as the mass of species $Z$ in the ISM divided by the total mass $M_G$ of ISM gas and dust. The concentration $Z$, rather than total numbers of atoms, is the quantity traditionally used in chemical evolution studies because it is concentration that astronomers measure. The mass $m_Z = Z M_G$ of interstellar species $Z$ is governed by					
\begin{equation}\label{eq13}
\frac{dm_Z}{dt} = -Z \psi(t) +Z_E \mbox{ } E(t) + Z_f \mbox{ } f(t)
\end{equation}

\noindent where $Z_E$ and $Z_f$ are, respectively, the concentration of Z in the spectrum of stellar ejecta, where it is large, and in the infalling gas where it is small. The metallicity in infalling gas $Z_f$ may probably be neglected with good accuracy.

	Many workers have shown that it is a good approximation for analytic understanding to assume that the ejecta $E(t)$ is returned at once from the entire spectrum of newly born stars despite the dependence of stellar lifetime on stellar mass. That assumption is called the \index{instantaneous recycling approximation} \emph{instantaneous recycling approximation}, and it is reasonable except late in the life of a galaxy when the gas mass and the star-formation rate have both become very small. But for galaxies in early and middle lifetime, eq. \ref{eq13} can then be written after some straightforward algebra as
\begin{equation}\label{eq14}
\frac{dZ}{dt} = y \mbox{ } \omega - \frac{Zf(t)}{M_G(t)} - \lambda \mbox{ } Z
\end{equation}
where $m_Z =ZM_G$, $Z$ being the concentration in the ISM gas (taken to be well mixed), and $y$ is the \emph{yield} of element $Z$, defined as the mass $m_Z$ of new $Z$ ejected from the entire spectrum of newly born stars divided by the mass of stellar remnants left behind by that entire spectrum of stars. The yield $y$ for primary nucleosynthesis products may be taken as a constant despite having small variations in full numerical models. The product $y\omega$ in eq.\ref{eq14} can be called the \emph{galactic nucleosynthesis rate}, which, through $\omega$, depends explicitly on the star formation rate ($\omega M_G = (1-R)\psi(t)$). Equation \ref{eq14}  for the ISM concentration differs from eq. \ref{eq10}  for the history of solar nuclei by the existence of the second term of eq. \ref{eq14}. That term reflects the loss of interstellar metal concentration $Z$ when ISM gas containing $Z$ collapses into new stars and is simultaneously diluted by metal-poor infall.

	Comparison of eq.\ref{eq14}  with eq.\ref{eq10}  shows clearly why it is an error to integrate eq. \ref{eq10}  thinking that the solar-nuclei birthrate spectrum can be mentally equated with the spectral rate of galactic nucleosynthesis. Equation \ref{eq14}  similarly has the galactic nucleosynthesis rate as the first term and the radioactive decay rate as the last term; but the astrophysical eq. \ref{eq14}  contains the middle term involving both the galactic infall rate and the mass of ISM gas, both of which are time dependent. Integrations of eq.\ref{eq10}  by assuming $B_{\odot}(t)$ ignore the effects of the ISM and can be correct only if the assumed form $dB_{\odot}(t)/dt$ actually resembles the number of solar nuclei born per unit time rather than the galactic nucleosynthesis rate. As the birthdate of solar nuclei is usually what is being sought in nuclear cosmochronology, one cannot get the answer from eq. \ref{eq10}  without first knowing and inserting the answer. For that reason nuclear cosmochronology must instead be investigated within the context of galactic chemical evolution. These points are central to the subject of astronomy with radioactivity within solar-system nuclei.

	\citet{Clay88} showed one way forward. One can integrate eq.\ref{eq14} analytically with the aid of flexible families of functions that enable analytic integration. This can be accomplished by form-fitting the ratio $f(t)/M_G(t)$ to an integrable family of functions $d\theta/dt$. Such form fitting is much more flexible than it would at first seem to be, because most physically plausible time-dependent behavior can be approximated by a specific form fitting choice. One such useful choice has been called the \emph{Clayton Standard Model}. It takes
\begin{equation}\label{eq15}
\frac{d\theta}{dt}= f(t)/M_G(t)= \frac{k}{(t+\Delta)}
\end{equation}
\noindent
and its explicit functions are given in Appendix~A1 of this book.  

	For the astronomy of radioactivity, the great merit of this approach is exact analytic solutions for $Z(t)$ for both stable and radioactive isotopes while simultaneously yielding exact functional representations of the mass $M_G(t)$ of interstellar gas and of the infall rate $f(t)$. Understanding these analytic models greatly aids understanding the behavior of radioactivity within more general numerical models of galactic chemical evolution. The numerical approach is to place the evolution of the galaxy on a computer, taking into account the evolutionary lifetime of each star formed and the specific nucleosynthesis products to be ejected from each star \citep[e.g.][]{Tim95}. Although this approach is undoubtedly correct, it obscures theoretical understanding that can be seen more easily within analytic models. Furthermore, surveys of nuclear cosmochronology can more easily be carried out within analytic models.

	All well mixed models of galactic chemical evolution \index{chemical evolution!galactic} can at best yield only an average expectation for the ISM. The true ISM is inhomogeneous in space and the nucleosynthesis rate is sporadic in time rather than maintaining its steady average. Nonetheless, well mixed models, both analytic and numerical, are important in laying out the results that would be true for a rapidly mixed ISM and a smooth rate of galactic nucleosynthesis. One interprets the observations of radioactive abundances against the backdrop of that expectation.

	Another result of great importance for short-lived galactic radioactivity is best illustrated within Clayton's  standard model. For short-lived radioactivity, i.e. whenever $\lambda$ is small, the mean concentration in the galactic ISM is
\begin{equation}\label{eq21}
Z_{\lambda} = y \mbox{ } \omega \left( \lambda +\frac{k}{(t+\Delta)} \right)^{-1}
\end{equation}
From this equation and one for a stable isotope the abundance ratio of a short-lived radioactive nucleus to that of a stable primary isotope in the ISM is larger by the factor $(k+1)$ than one might estimate without taking galactic chemical evolution into account. As an example, consider the interstellar ratio of radioactive $^{26}$Al to stable $^{27}$Al.  The formula
\begin{equation}\label{eq22}
\frac{Z(^{26}\rm{Al})}{Z(^{27}\rm{Al})} = \frac{y(26)}{y(27)} \mbox{ } (k+1) \mbox{ } \frac{\tau_{26}}{t_{\odot}}
\end{equation}
can be derived from such a ratio \citep{Clay93}. This result is larger by the factor $(k+1)$ than an estimate using only Eq. \ref{eq8}, which had traditionally been used in oversimplified discussions of the steady-state amount of short-lived interstellar radioactivity. That oversimplification can be found in almost all published papers on short-lived interstellar radioactivity. The extra factor $(k+1)$, which is an effect of infall of low-metallicity gas, became important when a large mass of interstellar $^{26}$Al was detected by gamma-ray astronomy (see Chapter~7).


\subsection{Gamma-Ray Lines from Galactic Radioactivity}
\label{sec:2-2-2}

One thinks of at least three reasons that the idea of astronomically detecting galactic radioactivity did not occur until the 1960s. In the first place, MeV range gamma rays do not penetrate to the ground, but must be detected above the atmosphere. Secondly, detecting MeV gamma rays and measuring their energies is a quite difficult technology, especially so when the background of cosmic-ray induced events above the atmosphere is so large. Thirdly, even a back-of-the-envelope estimate of expected rate of arrival from a stellar source is discouragingly small. \index{gamma-ray lines}

\subsubsection*{The Rice University Program}
In 1964 a new aspect of astronomy with radioactivity arose. Robert C. Haymes \index{Haymes, R.C.} was hired by Rice University, and he spoke there with Donald Clayton about the possibility of sources of galactic radioactivity that he might seek with an active anticoincidence collimation for a NaI detector flown beneath a high-altitude balloon.  \citet{B2FH} had speculated (incorrectly) that the exponential 55-night decline of the luminosity of many Type Ia supernovae was the optical manifestation of the decline of the spontaneous-fission radioactive decay of $^{254}$Cf \index{isotopes!254Cf} in the ejecta of the supernova, which would quickly have become cold without some heating mechanism that had to decline with a roughly two-month halflife. The $^{254}$Cf nucleus would be synthesized by the $r$~process (assumed to occur in Type Ia supernovae) along with uranium and thorium. The large kinetic energy of its spontaneous-fission fragments would be converted to optical emission by being degraded by atomic collisions within the ejected gas. This was called \emph{the californium hypothesis}. But B$^2$FH had said nothing about gamma-ray lines. The novel excitement at Rice University was the new idea that concepts of nucleosynthesis could be tested directly if the associated gamma-ray lines from the $r$-process radioactivity could be detected on earth coming from supernova remnants. Haymes estimated rather optimistically that his detector could resolve lines having flux at earth greater than about 10$^{-4}$ cm$^{-2}$ s$^{-1}$. The first scientific paper written with that goal was soon published \citep{Clay65}. It evaluated the full spectrum of radioactivity by an $r$~process normalized to the yield proposed by the Cf hypothesis. They found several promising lines. For example, the strongest from a 900-year old Crab Nebula would be \index{isotopes!249Cf} $^{249}$Cf, presenting 10$^{-4}$cm$^{-2}$s$^{-1}$ gamma-ray lines having energy 0.39 MeV. It is no accident that the $^{249}$Cf halflife (351 yr) is of the same order as the age of the Crab Nebula. Given a broad range of halflives in an ensemble of nuclei, the one giving the largest rate of decay has mean life comparable to the age of the ensemble. It can be confirmed that in a remnant of age $T$ the rate of decay per initial nucleus is $\lambda$e$^{-\lambda T}$, which is easily shown to be maximal for that nucleus whose decay rate $\lambda=1/\tau=1/T$. A photograph (Fig.~2.1) from those Rice years is included.

\begin{figure}
\begin{center}
\includegraphics[width=\textwidth]{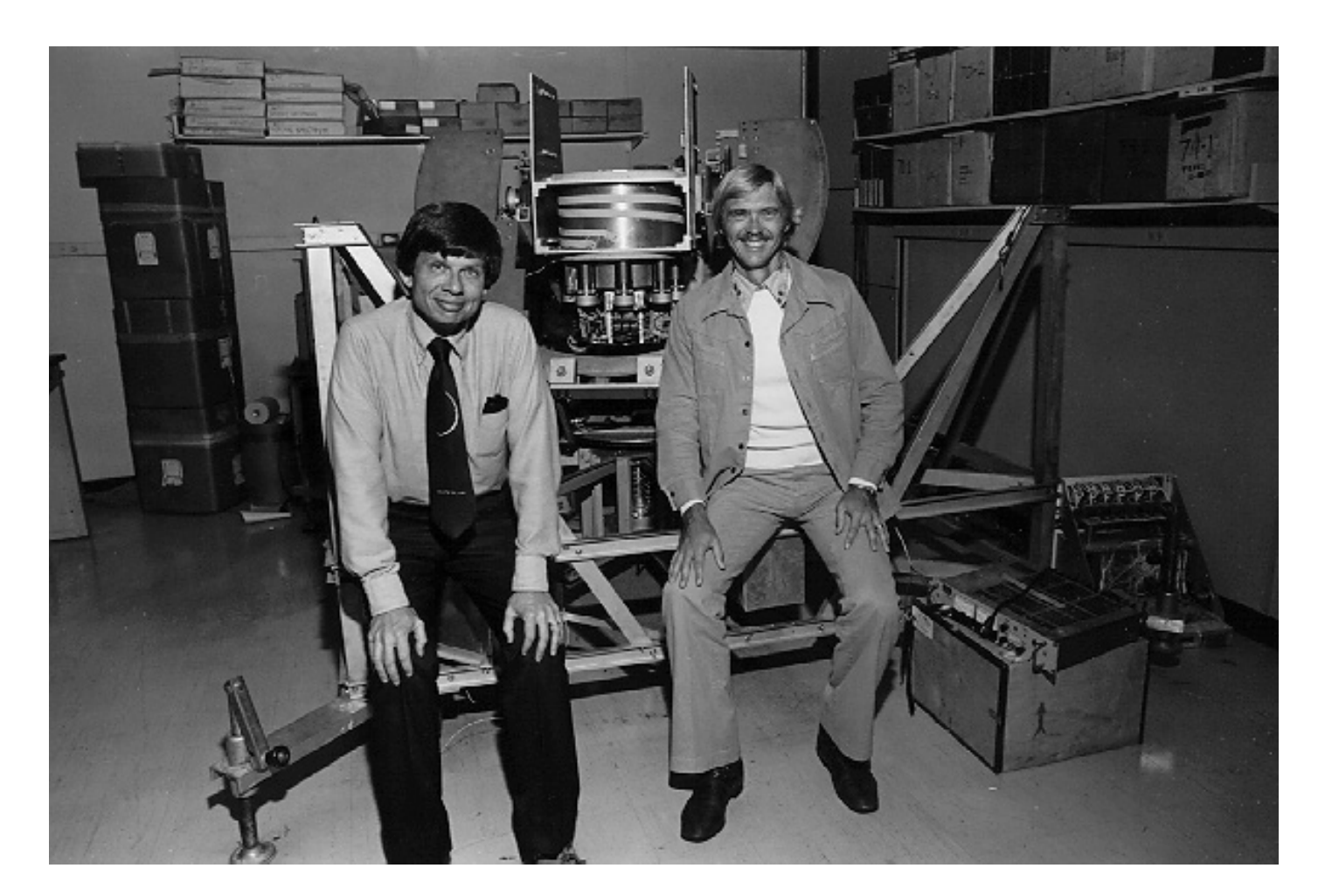}
\caption{Robert C. Haymes and Donald D. Clayton with Haymes' gondola for his balloon-borne gamma ray telescope at Rice University in 1973. Haymes \index{Haymes, R.C.} pioneered gamma-ray astronomy in the MeV region and Clayton developed gamma-ray-line targets for this goal and for nucleosynthesis. This program at Rice University trained two Principal Investigators and two Project Scientists for later experiments on NASA's Compton Gamma Ray Observatory. }
\end{center}
\label{fig_haymes}
\end{figure}

	Realistic expectations were not so sanguine, however. The $r$-process yield required by the Cf hypothesis appeared excessively large \citep{Clay65}. The $r$-process nuclei in solar abundances would be overproduced by a factor 100 if all Type Ia supernovae produced that yield. Despite that reservation, an exciting chord had been struck. A new astronomy of radioactivity appeared possible, one having significant scientific payoff, and Haymes's gamma-ray telescope began a program of balloon flights. The program did not discover $r$-process radioactivity, but it did discover apparent positron-annihilation radiation from the center of our galaxy \citep{John72} and hard x-rays from several sources. The observational program at Rice University also trained two of the principal investigators (G.J. Fishman and J.D. Kurfess) and two of the project scientists (W.N. Johnson and C. Meegan) on NASA's later {\it Compton Gamma Ray Observatory}, which went into orbit in 1991. And Clayton began a NASA-sponsored research program for laying out theoretical expectations for nucleosynthesis produced radioactivity and its associated nuclear gamma-ray lines in astronomy. Gamma-ray-line astronomy was in that sense launched by the Rice University program.

	At least two other groups followed the Rice University lead and became and remained leaders for the subsequent decades. They constructed differing telescope techniques from that of Haymes NaI scintillator. One was the JPL research group -- A.S. Jacobson, J. C. Ling, W. A. Wheaton, and W.A. Mahoney. \index{Jacobson, A.S.}\index{Wheaton, W.A}\index{Mahoney, W.A.} \index{Sch\"onfelder, V.} \index{von Ballmoos, P.} \index{Diehl, R.} Using a cluster of four cooled high-purity germanium detectors on NASA's {\it HEAO} 3, which was launched in 1979, they discovered the first galactic gamma-ray line from radioactive decay.  A group at MPI f{\"u}r Extraterrestrische Physik in Munich, V. Sch{\"o}nfelder, P. von Ballmoos and R. Diehl, developed a Compton-scattering telescope capable of good energy resolution and a greater angular view of the sky. Their active results from balloon-borne launches in the 1980s presaged the splendid results of their instrument on NASA's Compton Gamma Ray Observatory. \index{Compton Observatory} Jacobson's JPL group was not so lucky, as their experiment was removed from {\it CGRO} as a cost saving. Other groups joined the effort to create this new astronomy with radioactivity.

\subsubsection*{Earliest Predictions of Detectable Gamma-Ray Lines}
Inspired by the Rice University balloon program, a series of studies of the nucleosynthesis of radioactivity was undertaken. It may seem surprising in retrospect that this could not have been done by a routine search of the chart of nuclides in conjunction with nucleosynthesis theory. But the data base necessary did not exist. Four decades ago, nucleosynthesis theory was undergoing rapid expansion and clarification, so that it was a series of new insights that laid out each new prospect as it was discovered. Unquestionably the most important of these developments occurred in 1967 when a Caltech group \citep{Bod68A,Bod68B} and a Yale group \citep{Tru67} showed with differing research techniques that the $e$~process of \citet{Hoyle46} and  \citet{B2FH} for the nucleosynthesis of the iron abundance peak was incorrectly applied. Instead of producing iron isotopes as themselves within a neutron-rich nuclear-statistical equilibrium (NSE), as Hoyle had suggested, they were produced as isobars of radioactive nickel that decayed to iron isobars only after nucleosynthesis and ejection from supernovae. Owing to the increasing strength of the Coulomb energy in nuclei, the most stable nucleus for 4$n$ nuclei shifts from $Z=N$ below $A$=40 to $N=Z+4$ for $A$=44-60. But insufficient neutron excess required those major nuclei to form as unstable $Z=N$ nuclei. This brought into view the astronomical significance of several important radioactive nuclei. A historical study, \emph{Radiogenic Iron}, was written later to lay out the many significant astrophysical issues that hinged on this new understanding \citep{Clay93}.

It is of interest in a historical sense to consider the sequence of subsequent predictions of target radioactivity for astronomy. \citet{Clay69} jumped on the newly discovered radioactivity in the revised equilibrium processes (preceeding paragraph). Easily the most important nucleus was $^{56}$Ni, having $Z=N$, whose beta decays to $^{56}$Co and thereafter to $^{56}$Fe, having $N=Z+4$, are accompanied by numerous gamma ray lines that might be resolved by spectroscopic gamma-ray detectors. G. J. Fishman was a research student in R. C. Haymes' Rice University balloon program when he and Clayton  began calculating this spectrum of lines and their time dependences. Their motivation was to test nucleosynthesis theory, especially the fresh new view of explosive nucleosynthesis that had led to $^{56}$Ni nucleosynthesis. S. A. Colgate joined this effort because he had independently begun to investigate whether the energy deposition from the radioactivity could explain the exponentially declining Type Ia supernova light curves \citep{Colg69}. \citet{B2FH} had attributed those long lasting light curves to heating by spontaneous fission fragments from freshly produced $^{254}$Cf. As the simplest model \citep{Clay69} evaluated fluxes at earth from the rapid expansion of a solar mass of concrete, much of which had been explosively processed to $^{56}$Ni in the way anticipated by  \citet{Bod68A,Bod68B}. The lifetime of $^{56}$Co was seen as favorable for sufficient expansion for the gamma rays to escape. They neglected a massive envelope, as might overlie Type II cores, realizing that such an envelope would seriously limit the escape of the $^{56}$Ni \index{isotopes!56Ni} \index{gamma-ray lines} gamma rays. \citet{Clay69} was a very important paper, not only because it was correct in its prediction of the strongest gamma-ray nuclear source, but also because of its galvanizing effect upon experimental teams and on NASA planning. It later was included in the AAS Centennial Volume of seminal papers of the 20$^{\rm{th}}$ century. NASA Headquarters suggested funding of a research program at Rice to lay out additional prospects for this new astronomy of radioactivity. The NASA grant at Rice was entitled \emph{Prospects for Nuclear Gamma-Ray Astronomy}. But the detection of those $^{56}$Co gamma rays did not become possible until SN1987A occurred in 1987.

\citet{Clay69} also first suggested that unknown galactic supernovae (unseen owing to optical obscuration) may be discovered in gamma-ray lines. The lines from radioactive \index{isotopes!44Ti} $^{44}$Ti, with 60-yr halflife, ought to be detectable from several supernovae that have exploded during the past two centuries, considering that about three per century seem to occur on average within the Milky Way. This exciting idea has great implications for galactic nucleosynthesis and astrophysics; but its hope has been frustrated for astrophysical reasons that are not yet understood.

It was quickly realized \citep{Clay69B} that the entire universe might be filled with detectable $^{56}$Co \index{isotopes!56Co} gamma rays from the past history of supernova explosions throughout the universe. This work suggested that observable gamma rays owing to the collective effects of many supernovae rather than specific young supernovae might be targeted. Detecting the universe in this way remains a goal of the astronomy of radioactivity.

Two years later it was proposed \citep{Clay71} that the collective effects of galactic supernovae might allow one to observe gamma-ray lines from long-lived radioactive nuclei whose emission would be too weak from individual supernovae. That first work in that regard focused on \index{isotopes!60Fe} $^{60}$Fe, whose 2.6~Myr halflife\footnote{At that time, the halflife of $^{60}$Fe was best-known as 1.5~My} makes emission from an individual supernova too slow to be observable for the modest number of $^{60}$Fe nuclei produced within a single supernova. The $^{60}$Fe nucleus emits a 59 keV gamma ray upon decay, and its daughter $^{60}$Co emits gamma-ray lines of 1.17 and 1.33 MeV. Reasoning that during its long mean lifetime some 50,000 supernovae occur in the Milky Way, their collective effect should be observable. This reasoning applied equally well thirteen years later to the first interstellar radioactivity to be detected, that of $^{26}$Al. \index{isotopes!26Al} 

Surprisingly in retrospect, several years passed before \citet{Clay74} realized that when \index{isotopes!57Co} $^{57}$Co, the daughter of 36-hr $^{57}$Ni, decays, it also emits favorable gamma-ray lines and that its longer 272-day halflife would cause it to radiate these after the $^{56}$Co was almost extinct. This made $^{57}$Co much more significant than had been appreciated. \citet{Clay74} drew attention  to the significance of $^{57}$Co and to several new ideas for astronomy with radioactivity. Firstly, given an appropriate galactic supernova one might measure the material thickness overlying the radioactive cobalt by the distinct times at which differing gamma-ray lines reach maximum flux. This happens because the structural opacity for the 2.60 MeV gamma ray is only about half that for the 0.84 MeV gamma ray (both from $^{56}$Co decay), so that the 2.60 MeV gamma ray should peak earlier despite the equal rates of emission. Additionally, the 0.12 MeV gamma ray from the slower $^{57}$Co decay suffers even more absorption and so peaks later yet in time. These time delays have not yet been measurable because a time-dependent gamma line flux requires a bright supernova which would have to occur within our own galaxy. \citet{Clay74} also warned that hydrodynamic \emph{instabilities could allow the central material to burst through in streams} and thereby appear earlier than anticipated. Early escape of lines from $^{56}$Co was later detected in \index{supernova!SN1987A} SN1987A, which exploded in the Magellanic Clouds, and from which gamma-ray lines from both $^{56}$Co and $^{57}$Co were first detected (see Chapter 3). Earliest of the recorded $^{56}$Co photons were by the gamma ray spectrometer aboard NASA's {\it Solar Maximum Mission}, which happened to be in orbit when SN1987A exploded and whose sun-pointing spectrometer was reinterpreted as a gamma ray telescope \citep{Lei90}. Several balloon-borne gamma-ray spectrometers were flown and also detected $^{56}$Co lines. When $^{57}$Co was detected in SN1987A by \citet{Kur92} with the {\it Oriented Scintillation Spectrometer Experiment} on NASA's {\it Compton Gamma Ray Observatory} \index{Compton Observatory!OSSE}  during summer 1991, four years (about six halflives of $^{57}$Co) had elapsed since the explosion, so that the $^{57}$Co abundance had decayed to only about 1.5\% of its initial amount. Tension surrounded the hope that it would still be detectable. It was, with 4$\sigma$ significance, implying that the initial abundance ratio $^{57}$Ni/$^{56}$Ni was about twice the ratio $^{57}$Fe/$^{56}$Fe measured in terrestrial iron.

At almost the same time as the $^{57}$Co prediction, \citet{Clay74B} proposed that gamma rays may be detectable from the more common nova explosions owing to the radioactivity created by the nova outburst. These involve the positron-annihilation line from the hot-CNO burning that powers the nova outburst \citep{Star72}. These 511 keV photons would have to be detected very quickly, within roughly 10$^3$ s after the outburst because of the short halflives of CNO radioactive nuclei. They also pointed out the prospect of detecting a 1.274 MeV gamma ray following $^{22}$Na decay. Because of its longer halflife (2.6 yr), $^{22}$Na \index{isotopes!22Na} remains alive for the roughly 10$^6$ s required for the nova ejecta to become transparent to the gamma ray. These goals have not succeeded yet, but they remain a realistic hope of measuring the thermonuclear power of the nova. The model of the nova makes it a remarkable laboratory for thermonuclear explosions. Detection could confirm the model or rule it out. Some years later the possibility arose that the 478 keV gamma-ray line following the decay of radioactive $^7$Be to $^7$Li \index{isotopes!7Be} might also be detectable from novae \citep{Clay81}. Detectabilty requires the nova envelope accreted from the companion star to be enriched by an order of magnitude in $^3$He, but that is perhaps to be expected. If so, the nuclear reaction $^3$He $+$ $^4$He $\Rightarrow$ $^7$Be $+$ $\gamma$ creates the radioactive nucleus in the thermonuclear flash. One positive aspect is that the $^7$Be halflife is large enough for the envelope to become transparent.

 The radioactive $^{22}$Na nucleus is also produced in supernovae, although the complicated details of its nucleosynthesis therein were not well understood when \citet{Clay75C} proposed its detectability within supernovae. It might be observable for a decade, allowing ample time to become transparent to the $^{22}$Na gamma-ray line.

Almost the last good prospect to be predicted turned out to be the first actually observed! That was the radioactive $^{26}$Al \index{isotopes!26Al} nucleus \citep{Ram77}. Like radioactive $^{60}$Fe, the $^{26}$Al nucleus is long-lived and thus decays too slowly to be detectable from individual explosions. It is the cumulative yield of many events over the past Myr or so that was discovered. This is the topic of the next section.

The predictions whose history has been recounted here placed tantalizing targets before the community of experimental physicists. Many of these assembled teams to design, build and fly gamma-ray spectrometers capable of detecting these lines. NASA entertained a grants program pursuing these goals, which create a new wavelength range for astronomy. Moreover, the will to fund the {\it Compton Gamma Ray Observatory} was strengthened by these specific hopes. That hope was fueled also by the surprising discovery of a gamma-ray line from $^{26}$Al nuclei in the interstellar medium.


\subsubsection*{The Surprise $^{26}$Al}

The first detection of an interstellar radioactive nucleus came as a surprise. The history of this radioactive isotope in astrophysics reveals that correct predictions can be made on the basis of inadequate reasoning, that scientists may be blindsided by their own excessive faith in their own pictures and beliefs, and that experimental discovery is the arbiter. The possibility of detecting $^{26}$Al nuclei by observing the 1.809 MeV gamma-ray line emitted following its decay to $^{26}$Mg was suggested by \citet{Ram77} and independently by \citet{Arn77}. Their interesting papers were not quantitative predictions because the grounds for their suggestions were not convincing. They suggested wrongly that interstellar $^{26}$Al nuclei would be detectable if the production ratio in massive stars is $P(^{26}$Al)/$P(^{26}$Mg) = 10$^{-3}$. However, that production ratio, which was expected by carbon burning in massive stars, was inadequate for detectability. The argument showing that that production ratio was unobservable \citep{Clay84} was submitted for publication only after interstellar $^{26}$Al was discovered \citep{Mah82,Mah84}. The predicted 1.809 MeV flux was too small by a wide margin to be detected by contemporary gamma-ray spectrometers. So something was wrong.

Using the gamma-ray spectrometer on NASA's third {\it High Energy Astrophysical Observatory} (HEAO 3)  \citet{Mah82,Mah84} reported a measured flux of 1.809 gamma rays that required about 3M$_{\odot}$ of interstellar $^{26}$Al nuclei. \index{High Energy Astrophysics Observatory (HEAO)} That quantity was far larger than the predicted interstellar mass. Nonetheless, both \citet{Ram77} and \citet{Arn77} had urged, on general grounds and on intuitive arguments, that the 1.809 MeV interstellar gamma-ray line be sought. It was indeed observable, the first to be detected.

What were the conflicts that this discovery illuminated? Using Eq. \ref{eq8} for the number of radioactive nuclei born during the last mean lifetime ($\tau$= 1.05$\times$10$^6$yr) of $^{26}$Al nuclei, the time-average interstellar abundance ratio for aluminum isotopes was traditionally estimated to be
$^{26}$Al/$^{27}$Al = $P(^{26}$Al)/$P(^{27}$Al) $\tau$/10$^{10}$yr = 10$^{-7}$ if one uses $P(^{26}$Al)/$P(^{27}$Al)=10$^{-3}$. Since total ISM mass is about 10$^{10} M_{\odot}$, it contains about 5.8$\times$10$^5 M_{\odot}$ of stable $^{27}$Al. Multiplying by the above isotope ratio, the ISM would carry about 0.06 M$_{\odot}$ of $^{26}$Al nuclei, woefully inadequate for the observed 3M$_{\odot}$ of interstellar $^{26}$Al \citep{Mah82,Mah84}. So although \citet{Ram77} and \citet{Arn77} correctly suggested that $^{26}$Al nuclei might provide a suitable ISM radioactivity to seek, their reasoning did not convincingly justify that hope. The discovery of so much interstellar $^{26}$Al was surprising and meant that some assumptions were quite in error. In fact, several were in error.

Confirming detections of the $^{26}$Al radioactivity were quickly made from balloon-borne gamma-ray spectrometers \citep{Von87,Mac87}, and the total mass was later measured accurately by the gamma ray spectrometer aboard NASA's {\it Solar Maximum Mission}, whose sun-pointing spectrometer had been reinterpreted as a gamma ray telescope \citep{Share85}. The existence of 2-3 M$_{\odot}$, of $^{26}$Al was then beyond doubt. The 1991 launch of NASA's {\it Compton Gamma Ray Observatory} with its imaging COMPTEL Compton telescope (see Chap.10) \index{Compton Observatory!COMPTEL}  produced the most detailed data about the spatial distribution of interstellar $^{26}$Al \citep{Diehl94}.

Stepping back to the history of the mid 1970s, a related issue for the astronomy of $^{26}$Al radioactivity had arisen about one year prior to \index{Ramaty, R.} \index{Lingenfelter, R.E.} Ramaty \& Lingenfelter's 1977 suggestion; namely, it was discovered \citep{Gray74,Lee77} that the molecular cloud from which the sun had formed apparently carried within it the large isotopic ratio $^{26}$Al/$^{27}$Al=5$\times$10$^{-5}$ at the time the \index{solar system} planetary system began to form. Had that ratio been applicable to the ISM as a whole, it would have corresponded to 29 M$_{\odot}$ of radioactive $^{26}$Al nuclei-very much more than was discovered in 1982 by Mahoney {\it et al.} The seeming impossibility of producing such a large quantity by nucleosynthesis prompted the idea \citep{Cam77,Wass82} that a supernova within the molecular cloud from which the sun was born had injected the solar $^{26}$Al radioactive nuclei into the collapsing solar portion of the cloud just prior to the solar birth. In that case the solar cloud was quite atypical of the ISM at large. The initial solar $^{26}$Al nuclei in meteorites became the standard bearer for a class of extinct radioactive nuclei that required special local production realted to solar birth. But owing to the huge consequent ratio in the forming sun, it seemed to bear no transparent relationship to the existence of several solar masses of interstellar $^{26}$Al nuclei, which necessarily represents nucleosynthesis of $^{26}$Al nuclei by many thousands of supernovae spread out in time over 1-2 Myr prior to today. This discovery from meteorites brought the idea of interstellar $^{26}$Al nuclei to the attention of astrophysicists.

What new ideas brought the expected 0.06M$_{\odot}$ of $^{26}$Al nuclei in the ISM into line with the observed 3M$_{\odot}$ of $^{26}$Al nuclei? Some suggested that novae \citep{Clay84,Woosley80} or AGB stars, rather than supernovae, were the source of the $^{26}$Al nuclei; but that possibility was ruled out by observations made later with the Compton Telescope (COMPTEL) following the launch of $Compton$ $Gamma$ $Ray$ $Observatory$. Those observations \citep{Diehl94} showed spatial concentrations of $^{26}$Al nuclei in ISM regions where star formation is currently active. Spatial correlation of $^{26}$Al nuclei with massive stars was clear. Attention therefore returned to supernovae sources with the realization \citep{Aro80} that hot hydrogen burning shells of massive stars constitutes a significant extra source of $^{26}$Al. The small production ratio in carbon burning ($P(^{26}$Al)/$P(^{27}$Al) =10$^{-3}$) must be augmented by $^{26}$Al-rich convective shells in massive stars. Those calculations became a major industry with numerical models of presupernova evolution. Because $^{26}$Al/$^{27}$Al ratios of order 0.1 exist in these shells, their contribution calculated by using time-dependent numerical models of presupernova evolution would be needed.  \citet{Wea93} calculated that the average production ratio from a standard spectrum of massive stars was near P($^{26}$Al)/P($^{27}$Al)=0.006, six times larger than the earlier estimates. Secondly, the estimated ratio of mean interstellar $^{26}$Al/$^{27}$Al must be increased because most of the stable $^{27}$Al is locked up inside old stars whereas the live $^{26}$Al nuclei will still be overwhelmingly in the gaseous ISM. This reasoning augments the expected ratio by the factor (k+1) derived from the standard analytic models of chemical evolution of the mean galaxy. \citet{Clay93} presented that argument for the expected interstellar ($^{26}$Al/$^{27}$Al) abundance ratio. Because k=2-4 seems likely from other astrophysical arguments, this (k+1) correction also increases the interstellar ratio by another factor 3-5. These two effects combined amounted to an increase of a factor 20-30 in the initial expectations, bringing the expected value from 0.06 $M_{\odot}$ to 1-2 $M_{\odot}$ of $^{26}$Al nuclei in the ISM. The original conflict with global theory was largely resolved.

	Moreover, the largest observed flux from $^{26}$Al nuclei concentrated in ISM regions where star formation is currently active. This directly challenged the theoretical simplification that adopts a star-formation rate that is constant in time and spread smoothly through the spatial ISM. Models of chemical evolution of the galaxy use that simplification in order to be calculable. But the observations of 1.809 MeV gamma rays painted a picture of star formation occurring sporadically whenever and wherever large regions of ISM become vulnerable to prolific star formation. The massive $^{26}$Al emission regions stunned and surprised many experts (including this writer). The fluxes from these are moderated by the inverse square of their respective distances, further complicating discussions of the mass of interstellar $^{26}$Al.
	The history of $^{26}$Al nuclei in the ISM provides a textbook example of scientific progress into a new discipline. Experts disagree in their predictions; correct predictions often rely on intuition as much as on scientific justification; surprises often attend exploratory experimental surveys; and scientists from many disciplines amplify the relevant ideas and measurements. Astronomy with radioactivity is such a science, and its first discovery of an interstellar gamma-ray line emitter, the surprise of $^{26}$Al, was bathed in confusion.

\subsection{Radiogenic Luminosity}
\label{sec:2-2-3}
A big problem lay in the path of trying to understand how explosive objects in astronomy could remain bright. Their luminosity was expected to fall rapidly as the objects expand. Suppose a rapidly expanding object consists of an ideal nondegenerate gas and that the expansion is too rapid for gain or loss of heat; that is, the expansion is adiabatic. The internal heat of such an object is rapidly lost to the mechanical work of the expansion. For adiabatic expansion $TV^{\gamma-1}$ = constant during the expansion, where $\gamma$ is the ratio of specific heats at constant pressure $c_P$ to the specific heat at constant volume $c_V$. For a perfect monatomic gas $\gamma$= $c_P/c_V =5/3$. Thus $TV^{2/3}$ = constant for such an expanding hot object. If the expanding object can escape gravitational binding, the radius of an exploding object may be approximately $R=vt$, so that volume $V$ is proportional to $t^3$. Thus one would expect that $Tt^2$ is constant during the expansion. The problem is then that if luminosity $L$ is proportional to $R^2T^4$, as in a black body, one expects $L$ to be proportional to $t^{-6}$. However, many supernovae that are bright after one week are still bright after three weeks instead of dimmer by the large expected factor near $3^{-6}$. Similar problems exist for novae expansions. The problem, then, lay in discerning what source of energy could keep the expanding objects hot. The answer is radioactivity. The radiogenic luminosity of exploding objects became one of the major aspects of astronomy with radioactivity. \index{radioactivity}

\subsubsection*{Radiogenic Luminosity in Supernovae}

It was excellent luck that SN 1987A exploded nearby in the Magellanic Cloud in 1987.  It became the most observed explosive astronomical event of all time. Astronomers recorded its emissions in every possible wavelength band, which proved necessary in exposing how radiogenic luminosity worked in that kind of supernova. The observations showed that after July 1987 the total power output of SN 1987A \index{supernova!SN1987A} declined exponentially for about two years,accurately tracking the 77.2-day halflife of $^{56}$Co. This confirmed that the energy of the positrons emitted and of the subsequent gamma-ray emission following each $^{56}$Co decay were efficiently converted to optical and infrared luminosity. This old idea for declining supernova light curves was first treated for the identification of $^{56}$Co as the relevant radioactivity by \citet{Colg69}. The observed value of the total luminosity of SN 1987A required 0.075$M_{\odot}$ of $^{56}$Ni \index{isotopes!56Ni}  to have been synthesized within the core matter that escaped the central neutron star. With each scattering of a gamma ray it loses roughly half of its energy to the recoil electron, which is quickly degraded into heat. Once its energy has been scattered below 5 keV the large photoelectric opacity converts its entire remaining energy to electron energy, which is also degraded into heat.

Supernovae come in two main types, with additional structural subtypes. Type Ia and Type II differ in how compact they are, how much radioactivity is produced, how that radioactivity is distributed, and how much envelope overlies the radioactivity. The nucleosynthesis of $^{56}$Ni is easily the largest and most important source of heating power for the radiating gas. Such issues impact the way in which radioactivity provides luminosity for the remnant. To truly judge the degree of scientific understanding of radiogenic luminosity requires astronomers to infer, to the best of their ability, the structure of the exploding object. The luminosity being observed has complicated relationship to the location of the radioactivity. In SN1987A itself, for example, the presupernova star was blue, implying smaller size than the common red presupernova stars. The amount of overlying matter was modest owing to a large extent of mass loss during the prior evolution of the star. These characteristics influenced the relative importance of the radiogenic luminosity to the luminosity caused by the shock heated envelope.

Although $^{56}$Ni provides the largest and most important source of heating power, other radioactive nuclei also play very important roles. The $^{56}$Co daughter of $^{56}$Ni is actually more observable in its effect on radioactive luminosity. This is primarily because its 77.2-day halflife allows it to remain alive after the 6.08-day parent $^{56}$Ni has decayed. Therefore, $^{56}$Co delivers heating power at a later time when the expanding remnant is much larger and has suffered more adiabatic cooling. In these circumstances, quantitative astronomy with radioactivity requires careful evaluation of how, when and where the energy released by radioactivity is deposited in the expanding gas \citep{The90}.

The nucleosynthesis of $^{57}$Ni \index{isotopes!57Co} also plays a role in radiogenic luminosity because it decays rather quickly to 272-day $^{57}$Co. Owing to its longer halflife, $^{57}$Co is still providing radioactive power when the $^{56}$Co has decayed to negligible abundance. This transition of dominant radioactive power occurs after about two years. When {\it Compton Gamma Ray Observatory} \citep{Kur92} \index{Compton Observatory!OSSE} measured with OSSE the actual mass of $^{57}$Co in SN1987A, it corresponded to twice the solar ratio $^{57}$Fe/$^{56}$Fe. This surprised many who had inferred its abundance to be five times the solar ratio, a large isotopic ratio that had been deduced on the basis of the bolometric luminosity exceeding the instantaneous power from $^{56}$Co decay \citep{Sun92,Dwek92}. The radiogenic luminosity truly deriving from $^{57}$Co decay was demonstrated  by the OSSE measurement to have been overestimated by a factor near 2.5. That discovery team \citep{Clay92} then advanced a new theoretical aspect of radiogenic luminosity, what they called \emph{delayed power}. They presented a model showing that when the remnant gets sufficiently dilute owing to its expansion, the rate of electronic recombination cannot keep up with the rate of ionization that had been produced somewhat earlier by residual amounts of $^{56}$Co radioactivity. Because of the high degree of ionization, the radiogenic luminosity from $^{56}$Co radioactivity begins to exceed the instantaneous rate of energy deposition from $^{56}$Co decay. In effect, the radiogenic luminosity reflects the rate of $^{56}$Co decay from a somewhat earlier time than the time of observation. This observed upturn in luminosity had been attributed to $^{57}$Co instead of delayed power from $^{56}$Co decay. Such time-dependence became a new aspect of astronomy with radioactivity.

A more general aspect of radiogenic emission lies in any astronomical investigation in which the gas possesses a higher degree of energy excitation than could be expected in the absence of radioactivity. For example, the hydrogenic lines of He$^+$ in SN 1987A were interpreted as a consequence of the high-energy photons escaping from the supernova interior. Hard X rays emanting from the expanding supernova 1987A were another clear example of Compton-scattered radioactivity gamma rays escaping from the interior; and only radioactivity can produce so many hard X rays. Another example might be the detectable presence of doubly ionized species, an ionization state that would make no sense in a purely thermal setting. All such radiogenic possibilities are aspects of radiogenic luminosity, and as such a part of astronomy with radioactivity.

A similar transition between identities of powering radioactive nuclei occurs when the decay power of \index{isotopes!44Ti} 60-yr $^{44}$Ti exceeds that of $^{57}$Co. And no doubt there exists a similar problem, namely, does the rate of radiant emission of $^{57}$Co power keep up with the rate of $^{57}$Co decay power. The radioactive $^{44}$Ti is created in almost the same location as are the Ni isotopes. This occurs in the explosive burning of oxygen and silicon \citep{Bod68A,Bod68B,Woosley73} and reflects the extent to which $^{28}$Si has been decomposed into the silicon-burning quasiequilibrium. But radiogenic luminosity from this $^{44}$Ti nucleus depends even more strongly on how its emissions are converted to luminosity and the extent of the time lag between radioactive decay and the creation of observed photons. Its contributions to delayed radiogenic luminosity require careful assessment of those issues.

Such considerations are significant in the scientific understanding of astronomy with radioactivity. The luminosity of a young supernova remnant is one of the easiest observations of radioactivity in astronomy; but its interpretation requires fairly accurate description of the structure of the exploding object.

\subsubsection*{Radiogenic Luminosity in Novae}

	The brightening of a nova explosion and the associated ejection of matter from its expanded envelope are also issues that are dependent upon the deposition of radioactive heat to the envelope \citep{Star72}. Those authors showed that a successful model needs to mix carbon from the surface of the underlying white dwarf into the hydrogen-rich accreted envelope in order to have sufficient radioactivity produced by the thermonuclear flash that triggers the nova event. That burning is the hot CN cycle.  They showed that radioactive decay keeps the envelope hot while it expands to larger radiating surface area and therefore increased luminosity. They also showed that without the \emph{radioactive afterburner}, mass ejection would not occur. \citet{Tru82} gives more nuclear-physics details of the nova explosions.

\subsection{Extinct Radioactivity and Immediate Presolar Nucleosynthesis}
\label{sec:2-2-4}
\subsubsection*{Xenology Revisited}
John Reynolds \index{Reynolds, J.} had discovered in 1959 that extinct radioactive $^{129}$I appeared to have existed initially in meteorites. The observed ratio to stable I is near $^{129}$I/$^{127}$I = 10$^{-4}$. This stood for decades as a nearly unique example. Then in the mid-1970s new discoveries and ideas greatly enlarged the context of astronomy with radioactivity. Using Eq. \ref{eq22} for the number of radioactive nuclei born during the last mean lifetime ($\tau$= 23.5$\times$10$^6$ yr) of $^{129}$I nuclei, the mean interstellar abundance ratio for iodine isotopes is expected to be \index{isotopes!129I} \index{isotopes!127I}

\begin{center}
$^{129}\rm{I}/^{127}\rm{I}$ $= P(^{129}\rm{I})$/$P(^{127}\rm{I})(k+1) \tau_{129}/t_{\odot} = 0.013-0.022 $
\end{center}

\noindent using $P(^{129}$I$)/P(^{127}$I$)=1.5$, $t_{\odot}$=8~Gy for the presolar duration of nucleosynthesis, and $k=2-4$ within the standard model of radioactivity in galactic chemical evolution. The measured ratio is very much smaller than this expected ratio. The interpretation was that the solar molecular cloud sat dormant, with no new nucleosynthesis input, for 4-6 mean lifetimes of $^{129}$I to allow decay to reduce its activity to the measured level. As plausible as this \emph{waiting period} seemed to be for the next two decades, it was not correct.

	Reynolds' laboratory also discovered that that extinct radioactive $^{244}$Pu \index{isotopes!244Pu} appeared to have existed initially in meteorites. It was measured by an anomalous pattern of several Xe isotopes that was consistent with the spontaneous-fission spectrum from $^{244}$Pu.  \citep{Hud89} presented more modern data setting its abundance relative to that of $^{232}$Th as $^{244}$Pu/$^{232}$Th  = $3\times10^{-3}$. The many studies of these first two extinct radioactivities came to be called \emph{xenology} because of their reliance on excess isotopic abundances of Xe isotopes in meteorites. With a production ratio $P(^{244}$Pu/$^{232}$Th)  = 0.5 and mean lifetime $\tau_{244}= 115\times10^6$ yr, the expected interstellar ratio would be 0.02-0.03 -- larger by tenfold than the solar amount. But in this case a sequestering of the solar cloud to allow $^{129}$I nuclei to decay would not sufficiently reduce the $^{244}$Pu abundance. Trying to resolve the xenology puzzles occupied many meteoriticists for two decades.
Then three new developments altered xenology studies.

Firstly, existence of such $^{129}$Xe-rich \index{isotopes!129Xe} and fission-Xe-rich interstellar dust was predicted \citep{Clay75A} by the first argument that dust containing evidence of extinct radioactivity should condense within supernovae and influence the early solar system. This was an exciting new idea in the astronomy of radioactivity. However, it appears that chemical ways of forming meteoritic minerals from dust without loss of the carried xenon isotopes is too implausible; so this new idea was not supported for isotopes of Xe. But the idea of such supernova dust had been firmly implanted among the concepts of astronomy with radioactivity; moreover, the future may yet find evidence of fossil xenon. Several other extinct radioactive nuclei were predicted to exist within supernova dust \citep{Clay75A,Clay75}, and many of these are now demonstrated to exist.

Secondly, the $^{26}$Al extinct radioactivity was discovered in meteorites.  \index{isotopes!26Al} \index{meteorites} But its mean lifetime is too short ($\tau_{26}$= 1.05$\times$10$^6$yr) for any to survive the sequestering that xenology had suggested. This led to the second big change. The seeming impossibility of maintaining the large measured quantity of interstellar  $^{26}$Al nuclei by nucleosynthesis prompted the idea that a supernova within the molecular cloud from which the sun was born had injected the solar $^{26}$Al radioactive nuclei into the solar collapse just prior to the solar birth. Such a unique association would render the solar cloud atypical of the ISM at large \citep{Cam77,Wass82}. The substantial solar $^{26}$Al abundance initially in meteorites became the standard bearer for a class of extinct radioactive nuclei that required special local production.  \citet{Cam77} suggested that far from being a coincidence, the supernova producing the solar $^{26}$Al nuclei also triggered the collapse of the solar molecular cloud by the overpressure its shock wave brought to bear on that cloud. That model for live short-lived radioactivity in the solar cloud came to be referred to as \emph{the supernova trigger}. But Cameron \& Truran went too far. They supposed that the xenology-producing radioactivities ($^{129}$I and $^{244}$Pu) were also injected by that supernova trigger. They ignored that the ISM already was expected to contain a hundred times too much $^{129}$Xe for Reynolds' measurement, not too little, and ten times too much $^{244}$Pu. So injecting those radioactivities seems to not be correct either. Nor does the idea of fossil $^{26}$Mg nuclei from Al-rich interstellar dust rather than live $^{26}$Al in meteorites seem correct. But the puzzles of extinct radioactivity had come more sharply in focus, and these new ideas were instrumental for astronomy with radioactivity.

Thirdly, additional extinct radioactive nuclei were discovered. So diverse were their lifetimes that the idea of exponential decay from a starting abundance became increasingly untenable. Then another new idea of great importance for astronomy with radioactivity appeared.  \citet{Clay83} pointed out that the concept of a homogeneous ISM was at fault. He argued that supernova ejecta enter a hot phase of an ISM in which matter cycles on average among three phases, but that the sun was born from the cold molecular cloud phase. This distinction greatly modifies the expectation of exponential decay in the ISM and of extinct radioactivities in molecular clouds.  \citet{Huss09} provide a modern list of the many extinct  \index{radioactivity!extinct} radioactivity for which evidence exists.

The revisiting of xenology and these three new developments left extinct radioactivity research changed conceptually.

\subsubsection*{Neon-E}
A different noble gas, neon, provided an early but poorly understood evidence of stardust in the meteorites. As with xenology, the $^{20}$Ne/$^{22}$Ne isotopic ratio can be studied by incremental heating of a meteoritic sample in a mass spectrometer. David Black discovered that in the Ivuna meteorite the measured isotopic ratio $^{20}$Ne/$^{22}$Ne \index{isotopes!22Ne} \index{isotopes!20Ne} decreased precipitously when the sample heating reached 1000$^o$C, dropping to a ratio near three, much smaller than the common ratio near ten. Variation by a factor three in a trapped noble-gas isotopic component was unprecedented. This suggested that some unknown mineral releases its neon gas near 1000$^{\circ}$C, and that that mineral contains $^{22}$Ne-rich neon.  \citet{Black72} suggested that this small $^{20}$Ne/$^{22}$Ne isotopic ratio was so bizarre that it can not be accounted for by conventional means within an initially homogeneous, gaseous solar system. Black suggested that interstellar grains that had formed somewhere where the $^{20}$Ne/$^{22}$Ne isotopic ratio was smaller than three had survived the origin of the solar system and imprinted the 1000$^{\circ}$C temperature outgassing with its isotopic signature. He called this exotic neon component Ne-E. This was a radical far-reaching conclusion, perhaps the first of its kind based on sound analyzed data rather than on pure speculation.

	But where might this dust have formed? \citet{Aro74} remarked that neon gas ejected from explosive He-burning shells of massive stars could resemble Ne-E, but they offered no suggestion for forming carriers of Ne-E there or of otherwise getting such gas into the early solar system.  \citet{Clay75} offered a more concrete proposal, arguing that supernova dust would condense before the ejecta could mix with circumstellar matter, roughly within the first year, and that $^{22}$Na (2.6 yr) produced by the explosion would condense as the element sodium and decay to daughter $^{22}$Ne only after the grains had been grown.  In that manner nearly isotopically pure $^{22}$Ne could be carried within supernova dust into the forming solar system. This was an early prediction of the supernova-dust phenomenon in meteorites. \citet{Clay76} argued that nova dust provided another possible source for Ne-E.  Either radioactive venue would amount to a new technique for astronomy with radioactivity.

	The leading alternative to radiogenic $^{22}$Ne is condensation of dust in a $^{22}$Ne-rich red-giant stellar atmosphere. This can indeed occur, because during He burning the $^{14}$N \index{isotopes!14N}  residue of the prior CN cycle can be converted to $^{22}$Ne by two successive radiative alpha-particle captures \citep{Aro78}. The question arises whether this Ne is sufficiently $^{22}$Ne-rich. And that required much more study of meteoritic specimens. But the possibility was experimentally strengthened by the discovery \citep{Srin78} that at least one Ne-E component was associated with \index{process!s process} $s$-process xenon from red-giant atmospheres.

	These were exciting developments in isotopic astronomy and for astronomy with radioactivity. This fresh new set of ideas intensified interrelationships among interstellar stardust, the origin of the solar system, and meteorites.

\subsubsection*{$^{26}$Al: Fossil or Injected Fuel}
During the mid 1970s it was discovered \citep{Gray74,Lee77} that very refractory Al-rich minerals that apparently were among the first solids to form in the early solar system contained elevated isotopic ratios $^{26}$Mg/$^{27}$Mg. \index{meteorites} \index{isotopes!26Al} Furthermore, the number of excess $^{26}$Mg atoms within a mineral was shown to be proportional to the number of Al atoms. That correlation could be reproduced \citep{Lee77} if the Mg had been initially isotopically normal and the initial Al contained a small component of radioactive $^{26}$Al, whose later decay produced the excess $^{26}$Mg atoms.  The simplest way for this to have occurred is that the molecular cloud from which the sun had formed carried within it the large isotopic ratio $^{26}$Al/$^{27}$Al=5$\times$10$^{-5}$ at the time the planetary system began to form. The reason for calling that ratio large is that it far exceeded the ratio that had been anticipated to exist within the ISM. Had that ratio been applicable to the ISM as a whole, it would have corresponded to 29~\Msol of radioactive $^{26}$Al nuclei -- much more than was discovered in 1982 by Mahoney {\it et al.} by their detection of  the 1.809 MeV gamma ray line that is emitted following its decay.

The immediate issue was whether these radioactive $^{26}$Al nuclei were actually alive in the early solar system or whether $^{26}$Al was alive only when Al-rich dust formed in earlier galactic supernovae. If the latter option were true, the excess $^{26}$Mg nuclei were fossils of radioactive $^{26}$Al decay within interstellar Al-rich dust. Existence of such $^{26}$Mg-rich and Al-rich interstellar dust had been predicted \citep{Clay75A,Clay75} in the first works to propose that dust containing evidence of extinct radioactivity should condense within supernova ejecta. This was an exciting new idea in the astronomy of radioactivity. \citet{Clay77A,Clay77B} presented a solid-carrier model for excess $^{26}$Mg nuclei in Al-rich minerals based on the fossil picture. In the first case, that of live $^{26}$Al nuclei during the formation of the minerals, some explanation was required for why so much radioactive $^{26}$Al should have been present in the solar matter when it could not have been a general property of the ISM. In the second case, fossil excess $^{26}$Mg nuclei, why did interstellar Al dust containing abundant radioactive $^{26}$Al nuclei form within supernova ejecta and how did it participate in the chemical growth of the Al-rich minerals found in meteorites. In the first case, the heat of the radioactive decay of radioactive $^{26}$Al nuclei within those planetesimals that had been the parent bodies for the meteorites would have been sufficient to melt large parent bodies, allowing them to differentiate chemically (as the earth has done). In that case $^{26}$Al also played a role as fuel for melting of the meteorite parent bodies. In the second case, the excess $^{26}$Mg nuclei represented a fossil of interstellar decay. Tension between these two possibilities was reflected in the phrase \emph{``Fossil or Fuel''} within the title of the paper by Lee {\it et al.}, who argued in favor of the live $^{26}$Al nuclei in the solar gas. This debate raged for a few years but lost steam as chemical arguments for the growth of the mineral phases seemed increasingly likely to require live $^{26}$Al nuclei.

The struggle over the interpretation very short-lived extinct radioactivity was heightened later by the discovery of extinct $^{41}$Ca \index{isotopes!41Ca} in meteorites. Its abundance relative to that of $^{41}$K is only $^{41}$Ca/$^{41}$K = 1.5$\times$10$^{-8}$ \citep{Srin96}; however, even that small abundance looms large because its mean lifetime is but $\tau_{41}$= 0.144$\times$10$^6$ yr. Surely none can survive galactic nucleosynthesis to the time the solar system formed. So the supernova trigger injection would be needed to account for it as well as $^{26}$Al. But fossil effects are also definitely possible. Huge $^{41}$K excesses in Ca-rich supernova dust had been predicted two decades earlier \citep{Clay75}.

The history of documentation of extinct radioactive nuclei that had been alive in the early solar system was led for decades by G.J. Wasserburg \index{Wasserburg, G.J.} and his laboratory at Caltech (jokingly called ``the lunatic asylum'' owing to their research on lunar samples). His laboratory was of such renown for its measurements using traditional mass spectrometry with extreme sensitivity that Wasserburg was chosen recipient of the 1986 Crafoord Prize for geosciences. Similar mass spectrometric research of great consequence for this problem was conducted by G\"unter Lugmair in his laboratory at UCSD.

\subsubsection*{Extinct Radioactivity and Mixing of Interstellar Phases}
Xenology soon gave way to a host of newly discovered extinct radioactive nuclei. $^{129}$I and $^{244}$Pu \index{isotopes!244Pu} could seen to be members of a larger class of now extinct parents. A very special class contains those nuclei having mean lifetimes long enough that some of that radioactive abundance might survive from galactic nucleosynthesis but short enough that their survival live to the time of the solar system depends on the timing of interstellar mixing. That class of nuclei also includes $^{129}$I and $^{244}$Pu. Their mean lifetimes are, in ascending order: $\tau$($^{60}$Fe)= 2.16\footnote{The mean life of $^{60}$Fe is now known as 3.8~My}$\times$10$^6$ yr; $\tau$($^{53}$Mn)= 5.34$\times$10$^6$ yr; $\tau$($^{107}$Pd)= 9.38$\times$10$^6$ yr; $\tau$($^{182}$Hf)= 13.8$\times$10$^6$ yr; $\tau$($^{129}$I)= 23.5$\times$10$^6$ yr; $\tau$($^{244}$Pu)= 115$\times$10$^6$ yr; and $\tau$($^{146}$Sm)= 149$\times$10$^6$ yr. The relative abundances that they possess and the three-isotope correlation plots by which they are measured are discussed in Chapter~6. \index{isotopes!60Fe} \index{isotopes!53Mn} \index{isotopes!107Pd} \index{isotopes!146Sm} \index{isotopes!129I} \index{radioactivity!extinct}  What became abundantly clear is that simple exponential decay from a set of starting abundances is not consistent with the data. Something more complex is at work.

{\bf Four layers of interpretation}
The new data and new creative ideas led to a picture of cosmic radioactivity that is interpreted on four layers of complexity. Like the proverbial \emph{Russian doll}, one opens each layer only to reveal a more complex version inside. What are these four layers?

\begin{itemize}
\item[1. ]
{\it Mean Galactic Nucleosynthesis and the well-mixed ISM:}
Using Eq. \ref{eq22} for the mean interstellar ratio of a radioactive abundance to that of a reference isotope gives that mean abundance ratio to be
\begin{equation}\label{eq22a}
N(^AZ)/N(^{A*}Z) = P(A^Z)/P(^{A*}Z)\mbox{ } (k+1)\mbox{ } \tau_Z/t_{\odot}
\end{equation}
If the halflife of species $^AZ$ is long enough that the ISM can be considered to be well-mixed, it may be interpretable by the mean expectation of Eq. \ref{eq22a}. An example might be the uranium isotope ratio, since both have halflives of 10$^9$ yr or greater. The ISM probably mixes well during 10$^9$ yr.
\item[2. ]
 {\it Mean Galactic Nucleosynthesis and the ISM Phases:}
	If the lifetime $\tau_Z$ is of order 10$^8$yr or less, the abundances of that radioactive nucleus may differ in the different phases of the ISM. The three-phase mixing model \citep{Clay83} that describes this in the mean is described below. The idea is that supernova ejecta appear initially in the hot ISM, and considerable mean time is required for it to work its way via mass exchange between phases into the molecular clouds where the sun might have formed. That delay reduces radioactive abundances, depending upon their lifetimes. This idea regards each phase of the ISM as having the mean concentration of radioactive nuclei everywhere in that phase; but the three phases have differing mean concentrations of the radioactivity, with the molecular-cloud abundance being the least. This picture divides the mean ratio Eq. \ref{eq22} into differing steady-state ratios in each of the three phases. An example might be the iodine isotope ratio, since its halflife of 17~My is shorter than the expected phase change times (perhaps 50 Ma) for the ISM phases. Thus each phase will in the mean have different $^{129}$I/$^{127}$I ratios.
\item[3. ]
 {\it Mean Galactic Nucleosynthesis and Deviations from mean ISM mixing times:}
	All galactic samples of a given ISM phase will surely not be identical. \index{chemical evolution!galactic} Differences occur because nucleosynthesis does not follow the galactic mean rate at all locations, or because the interphase mixing times may vary from one place to another. Separate portions of a phase will scatter about that mean for that phase, perhaps by sizeable amounts.  Astronomical observations of very old, very metal-poor stars reveal this phenomenon clearly \citep{Burris00}. The iodine isotopes can illustrate how this works. Consider the molecular cloud from which the sun formed. It may have existed in a portion of the galaxy whose hot phase received less than the average amount of new nucleosynthesis during the past 24~My (the $^{129}$I mean lifetime). Thus the $^{129}$I concentration locally would be less than its galactic mean, whereas stable $^{127}$I would have the same concentration everywhere.  Furthermore, the time delay for the hot phase to be admixed into the solar molecular cloud may locally have exceeded the galactic mean delay, in which case the $^{129}$I concentration in the solar molecular cloud would have decayed to a smaller value than typical of that phase. This layer of interpretation deals with specific local galactic workings that differ from the galactic mean.
\item[4. ]
 {\it Nucleosynthesis from a single neighboring supernova:}
	The solar concentration of a short-lived radioactive nucleus may have been the result of a single galactic supernova. The shorter the halflife, the more likely it becomes that a neighboring dying star is needed to create it. In such a case the concept of mean galactic nucleosynthesis rate has no relevance. Short-lived $^{26}$Al is historically the prime example of this situation. Variations on the supernova injection model then appear: When did that supernova occur? With what efficiency did its ejecta admix into the solar molecular cloud? How long was required for that mixing to occur? Such questions are so specific that they are frustrating. \index{solar system} Almost any answer seems possible, depending upon hydrodynamic modeling for its credibility.
\end{itemize}
	Finally, some radioactive nuclei may have multiple causes for their solar abundances. Each layer of interpretation may contribute. Take $^{244}$Pu as a likely example ($\tau$($^{244}$Pu) = 115 Ma). A portion of the solar $^{244}$Pu abundance may survive at the level expected from mean galactic nucleosynthesis (layer 1). Layer 2 may have reduced that somewhat, depending on the mean interphase mixing times. Layer 3 describes a non-average $^{244}$Pu solar concentration owing to atypical (non-mean) local nucleosynthesis rate and mixing parameters. Finally, a portion of solar $^{244}$Pu abundance may have resulted from a single nearby supernova at the time of solar birth. It will be clear that scientific understanding  must seek the best picture for simultaneously fitting all of the radioactivities. Each individual abundance will be inadequate by itself to determine the physics of its solar presence. But a model that fits them all is a serious contender for the truth. To advance more into this topic, consider the mean interphase mixing model.

{\bf Mean mixing model for ISM radioactivity:}
Astrophysics provides this novel example of the layer-2 mathematics of radioactive decay. The distinct ISM phases have vastly differing temperatures and densities. These phases mix with one another on timescales that are longer than the shortest of the extinct radioactivities but shorter than the longest. As a consequence, the concentration $X$ of a radioactive nucleus will differ from phase to phase, with the differences greater for shorter halflife. Solar-system samples revealing the presence of once-live radioactive nuclei are obtained from meteorites and other planetary objects formed in the solar accretion disk. Measurements of solar samples therefore are of the concentrations in the cold molecular cloud (one of the phases) from which the sun was born. The key point is that radioactive concentrations in that molecular-cloud phase should be smaller than in the mean ISM, which includes all of its phases. This consequence of the mixing times among the phases, taking into account their levels of radioactivity, was devised by  \citet{Clay83}.

	{\bf A simplified two-phase model}  To focus on mathematical aspects of radioactivity rather than on aspects of astrophysics, consider first a simplified model that illustrates the essence of the problem. Suppose that the ISM consists of but two phases. Stars are born in phase 1, but the freshly synthesized radioactivity is ejected from supernovae into phase 2. Let those two phases have equal mass $M$ and exchange matter with the other phase on a timescale $T$. The mass exchange rate is then $(dM/dt)_{exch} = M/T$. Matter from phase 1 joins phase 2 at that rate, and conversely matter from phase 2 joins phase 1 at that same rate. The masses of each phase then remain constant and equal.

In reality, spatial inhomogeneity also exists, depending on where in phase 2 the fresh radioactivity is created and how long is required for its homogenization. Roughly 30,000 supernovae occur per My in the Milky Way, and they have similar nucleosynthesis yields, so an almost homogeneous injection phase occurs faster than the output of a single supernova can be spread uniformly. Even so, concentration differences and isotopic differences will exist spatially in that injection phase. The best that can be done without a specific calculation defined for a specific configuration is to calculate the \emph{average concentration} in the injection phase. To this end consider the injection phase to be uniformly mixed at all times (Layer 2). With that assumption the average difference in radioactive concentration between the phases can be calculated with a simple mathematical model. The answer can not be assumed to have applied exactly to the unique case of solar formation. Nonetheless, such a calculation reveals what is anticipated on average, without inhomogeneous effects.

Returning to the simplified two-phase model for those average concentrations, the mixing rate between them is $(dM/dt)_{exch}$ = $M/T$. Then with $X_1$ being the concentration (grams of $X$ per gram of ISM) of a radioactive nucleus in phase 1, the star-forming phase, its total radioactive mass $MX_1$ changes as follows:
$d/dt (M X_1) = M dX_1/dt + X_1 dM/dt = M dX_1/dt$
because the second equality is a consequence of the unchanging mass $M$ of each phase. How shall one evaluate $dX_1/dt$? It changes owing to two effects: first, the radioactive decay of $X_1$; second, the mass exchange with phase 2 having concentration $X_2$. One writes for those two terms
	$MdX_1/dt = -MX_1/\tau + (X_2- X_1)M/T$
where the first term is the radioactive decay rate with mean lifetime $\tau$. But the left-hand side must also vanish because $dX_1/dt=0$ in the steady state. This reasoning yields at once
	$X_1 = \tau/(T+ \tau) X_2$
This factor expressing the ratio of the concentrations can differ significantly from unity.
Suppose the mixing time $T$=100 Ma. Then for $^{129}$I, $X_1$ = (23.5Ma/123.5Ma)$X_2 = 0.19X_2$.
The $^{129}$I concentration in star-forming clouds is then fivefold smaller than in the injection
phase. This reduction is even greater for a nucleus having smaller mean lifetime. But for
very long lifetimes, $\tau/(T+\tau$) approaches unity, both phases having the same
concentration.

	Both $X_2$ and $X_1$ can be individually evaluated from the requirement that their
average (since the two phases have the same mass)  must equal the mean expected ISM
concentration for that nucleus (Eqs. \ref{eq21} and \ref{eq22}). Chapter 6 includes a table of the
known extinct radioactive nuclei and their abundances in the early solar system.

	The significance for differing ratios of extinct nuclei between the two phases is that those differences do
not reflect the simple expectation of exponential decay. The ratio $\tau/(T+\tau$) is very far
from exponential in dependence on lifetime $\tau$. \citet{Clay83} introduced that
effect when comparing the solar abundance levels of differing radioactive
nuclei. This simplified two-phase mean model illustrates that idea.

\begin{figure}
\begin{center}
\includegraphics[width=0.8\textwidth]{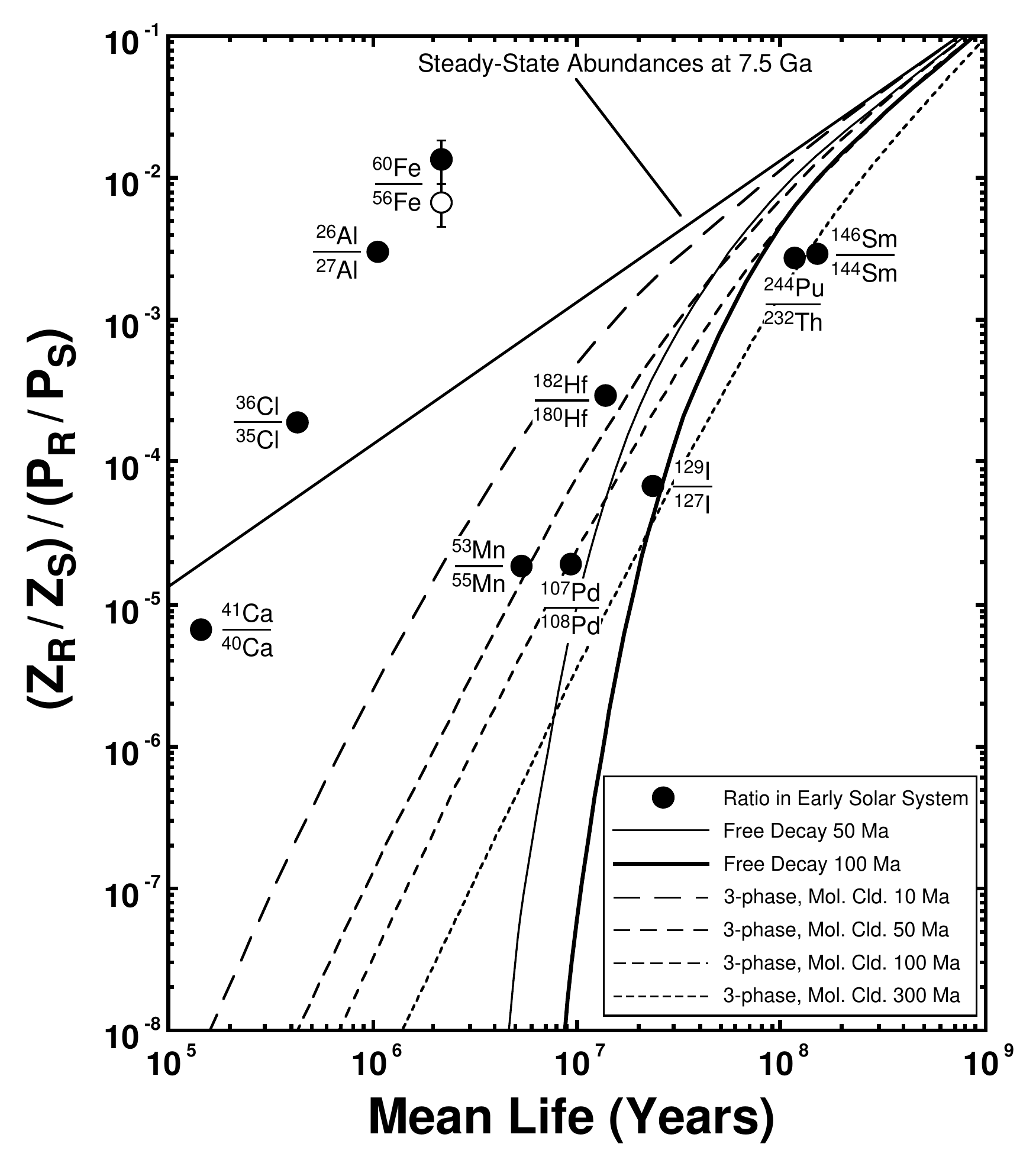}
\caption{Calculated surviving levels of the extinct radioactive nuclei in the solar cloud at the time of solar birth are expressed as a ratio to a neighboring stable isotope of the same element. Galactic steady state modefied by the continuous interphase mixing is shown for several different mixing times. \citep[from][]{Huss09} .}
\end{center}
\label{fig_huss}
\end{figure}

	{\bf Clayton three-phase model}  The initial \citep{Clay83} introduction of this physical idea
for interpreting relative abundances of extinct radioactive nuclei in this manner actually suggested a three-phase ISM rather than only two. Those masses and phases were:

(1)	$M_1$ is the molecular-cloud mass in which stars form. It exchanges mass with $M_2$.

(2) $M_2$ is the mass of large HI complexes that can not be disrupted by the supernova shock waves that frequently traverse the ISM. Phase 2 exchanges mass with $M_1$ with lifetime $T_1$ and also with $M_3$ with lifetime $T_2$.

(3) $M_3$ is the mass average HI clouds that are sufficiently small to be disrupted by passing supernova shock waves and which therefore are part of this warm neutral phase. Phase 3 exchanges mass with phase 2 with lifetime $T_2$.

It is not clear that this three-phase model is more realistic than the simpler two-phase model; but the need for fresh radioactivity to mix from its injection phase through a second phase before admixing with molecular clouds does amplify the distinction between radioactive lifetimes. Future progress with this multiphase idea, which definitely is significant for the extinct radioactivity problem, will require more modern astrophysical work on the differing phases, their masses, and how effectively they exchange mass with one another. Finally, it may be that time-dependence must be introduced. The growth of each phase, rather than static mean masses for each, needs evaluation. Over long times the global masses of each phase will be static; but their cyclic growth and disruption may be an essential part of the problem.

An insightful paper in regard to these issues has recently appeared \citep{Huss09}. Those authors present current data for the observed abundances of each extinct radioactivity. They also review the sites of nucleosynthesis of each nucleus. Then they evaluate the mean abundance expected within the standard model of galactic chemical evolution moderated by the interphase mixing described above. Fig.~2.2  \citep[from][]{Huss09} displays the expectation as a function of the mean decay lifetime for a wide range of interphase-mixing timescales and compares those results to those for two differing free-decay intervals (the solid curves). Note that interphase mixing fits more extinct radioactivities than does either free-decay interval;but the four short-lived isotopes are too abundant for any galactic scheme, apparently demanding injection from an adjacent supernova. Keep in mind that what is being examined by these mathematical procedures is a fundamental property of interstellar radioactivity, couched for convenience and transparency in terms of realistic analytic representations of the true galactic mixing processes.

\subsection{Stardust: Radioactivity in Solid Samples of Presolar Stars}
\label{sec:2-2-5}
A presolar stardust grain is composed only of atoms from a single donor star. The word {\it stardust} \index{stardust} applies to only those grains. It is used as a scientific word rather than a poetic word. It may be hard to accept that the isotopic composition of elements in an interstellar grain attests to it being a solid chunk of a single star. An average ISM grain has suffered a lengthy residence time within the interstellar medium before being incorporated into a new stellar accretion disk. Do not high energy interstellar ions implant within the grains as well as sputter them? Can interstellar chemistry add more atoms to mineralized grains? Would not grain ejection from a protostellar disk after chemically adding atoms to it within the hot dense disk structure also produce mineralized grains? Many questions flood the skeptical mind. Such doubts are reasonable considering the incomplete knowledge of these and other processes.

	Stardust grains are not average interstellar grains. Those grains themselves provide the best answers to the doubts expressed above. Their well ordered crystal structures suggest thermal condensation at high temperatures, as would be expected within hot but slowly cooling gases leaving a star. Annealed crystals would not, in contrast, be expected from accreted interstellar atoms. The dramatic isotopic ratios measured within individual stardust grains, obtained from nearly homogeneous grains having large numbers of atoms (10$^6$ to 10$^{12}$, say), strongly suggests that the condensation was from a gas having the extremely non-solar isotopic composition of the stardust grain.  The SiC grains, one of the most abundant of all presolar stardust grains, were assembled from C atoms having observed isotopic ratios in the grains between $^{12}$C/$^{13}$C=3 to 5000, whereas hot chemistry in an accretion disk would shift ratios toward interstellar norms, $^{12}$C/$^{13}$C=89 in the case of the solar accretion disk. Similar evidence appears in isotopic compositions of N and Si, to name only the most studied of the elements. Those ratios attest to an extreme condensation environment wherein isotopic composition is not that found in the ISM. These isotopic studies of individual stardust grains have been made using SIMS (secondary-ion mass spectrometry) in terrestrial laboratories \citep{Bern97B}.

What followed from the discovery of stardust was nothing less than a revolution in astronomy. \index{astronomy} \index{mass!spectroscopy} Solid pieces of stars are now characterized routinely in terrestrial laboratories, including isotopic analysis of many elements with a precision not attainable at telescopes. Distinguishing isotopes in stellar spectra is very difficult. Three scientists dominated the early development of this field experimentally. Robert M. Walker committed his laboratory at Washington University to development of new technology capable of better laboratory study of primitive solar system samples. He foresaw the capability of the sputtering ion probe, and recruited Ernst Zinner to devote a decade to improving and studying that technique. Their laboratory at Washington University (St. Louis) was then ready to study the newly isolated meteoritic grains, which they documented to be stardust. Zinner \index{Zinner, E.} and his many students led and dominated the new astronomy that blossomed from their isotopic analysis. An engaging history of this has been written in tribute to Walker and Zinner by \citet{McK07}. Thirdly, Edward Anders \index{Anders, E.} inspired his associates at University of Chicago to the detective-story pursuit that isolated stardust from the bulk meteorite rock. They termed their approach \emph{burning down the haystack to find the needle}. It tracked the chemical carriers within the meteorites \index{meteorites} of isotopically anomalous neon and xenon through a sequential dissolving the bulk of the rock in strong acids \citep{Amari94} and finding that the undissolved residue became increasingly anomalous. The carriers of anomalous noble gas were, fortunately, carbonaceous, and therefore did not dissolve in acid. From these residues they were able to extract the individual stardust grains. This too was a profound adventure in scientific exploration. Finally it may be noted that predictions of the existence of isotopically anomalous stardust had existed for a decade prior to their experimental discovery.

Within oxide grains \citep{Nitt97} stunningly abnormal families of correlated O isotopes rule out any growth within mixed ISM. Equally stunning was the almost pure isolated $s$-process isotopic compositions that were predicted for noble xenon in red giant stardust \citep{Clay78} and discovered \citep{Srin78} in bulk carbon-rich residues of acid dissolution of meteorites. These Xe$_s$-rich residues are dominated by collections of SiC grains. By contrast, interstellar bombardment would implant normal Xe isotopic composition. Neon isotopes also revealed an exciting story. Individual mainstream SiC grains have been shown to carry almost pure $s$-process \index{process!s process} Mo, a quite reactive trace element rather than a noble gas, showing that the Mo-containing condensing gases had not mixed with the ISM prior to condensation. Isotopic evidence of stardust abounds. By careful consideration of the entirety of the known properties of stardust grains, researchers become confident that they are indeed solid samples of stars that died before the sun was born.

\subsubsection*{Stardust predicted}
With the realization that computer models of massive-star supernovae generate shells of abundant intermediate-mass elements,  \citet{Hoyle70} argued that the adiabatic expansion and cooling of these newly synthesized nuclei within the supernova interior should be accompanied by condensation of a potentially large amount of silicate dust. The ejected gas cools to below 2000K after only several months when the density is still quite high. Their goal had been to account for the interstellar dust mass of silicates and of graphite. It did not occur to them that such grains could be identified by their unusual isotopic compositions. Nor did they suggest that they might be found within meteorites. They simply tried to explain the observed existence of dust in astronomy.

 \citet{Clay75A,Clay75,Clay78B} advanced exciting observational considerations on the grains' condensed radioactive nuclei. It was suggested that grains from the supernova-condensation process would be identifiable by their extreme isotopic signatures, inherited from the isotopic compositions of those supernova shells. Radioactivity is prominent within supernova shells. Short-lived radioactivity was proposed to decay within each grain during its interstellar residence. So the ISM should contain everywhere numerous interstellar grains from the very large number of presolar supernovae.  Also proposed was a chemical memory model \citep{Clay77A,Clay77B,Clay78B,Clay82} in which supernova grains were incorporated during later growth in the solar accretion disk of the larger solar-system solids. Many of those solids were incorporated into meteorites. That picture suggested potential explanations for several isotopic anomalies that had been discovered in meteorite solids; but it was rejected initially by meteoriticists, who focussed their doubts on the potential explanations of xenon anomalies from extinct radioactivity.  \citet{Clay76B} and \citet{Clay79,Koz89} demonstrated that a sufficient number of collisions of supernova-interior atoms will occur with any grain nucleation during expansion of the supernova interiors to grow grains greater than one micrometer in radius. None of these works was so optimistic as to predict that individual supernova grains would be found intact within meteorites; but they did establish the ideas by which those discoveries could be recognized. Radioactivity lay at the heart of these predictions.

In a paper outlining a system for the different types of precondensed matter in the \index{solar system} early solar system, the supernova condensates were named SUNOCONs and the term STARDUST was restricted to hot thermal condensation during mass loss from other stars \citep{Clay78B}, primarily red giants. NEBCON was suggested for grain mantles grown by nebular sticking of atoms and molecules to preexisting dust. These names have not found the favor of usage, despite describing a theoretical picture of what to expect in the early solar system. Only the term stardust is commonly used, and applied to all types of high-temperature thermal condensates from stars. The title \emph{Precondensed matter: key to the early solar system} of  \citet{Clay78B} explicitly contradicted that epoch's favored picture in which the initial solar system was initially hot and totally gaseous and condensed its solids from that solar gas.

The predictions of isotopically extreme stardust were, after some initial controversy, wonderfully confirmed by the discovery of supernova stardust and the later documentation of chemical memory within large solids,  earning broad acceptance.

\subsubsection*{Stardust discovered: Fossil Extinct Radioactivity}
Secondary-ion mass spectrometry (SIMS) \index{mass!spectroscopy} \index{SIMS} of ions sputtered from isolated single meteoritic grains by a focused ion beam identified stardust in the late 1980s. A history may be read in  \citet{And93}. First to be discovered \citep{Bern87,Zin89} was silicon carbide (SiC), with isotopic patterns that identified it as thermal condensation in matter flowing away from the photospheres of asymptotic-giant-branch red giants. This sensational identification of stardust is now beyond doubt, buttressed by a huge number of experimental and theoretical studies. These carbide grains can condense only after the composition changes caused by the third dredgeup of carbon, which eventually cause carbon to be more abundant than oxygen in the envelope. As long as O is more abundant than C, the CO molecule oxidizes (combusts) all carbon. Those SiC grains condensed after the carbon-star transition contained $^{26}$Al/$^{27}$Al near 10$^{-3}$ based on the excess $^{26}$Mg atoms within their magnesium-bearing phases. When supernova SiC $X$ grains were later identified as a subfamily of SiC stardust, many revealed $^{26}$Al/$^{27}$Al near 0.1 when they condensed within the expanding supernova interior. \index{isotopes!26Al} 

 Oxide stardust too was identified and classified into isotopic families a few years later \citep{Nitt97}. Its most common pattern is $^{17}$O-richness owing to secondary production in the parent star of $^{17}$O from initial $^{16}$O during hydrogen burning . At about the same time supernova stardust was identified unambiguously \citep{Hop96,Nitt96} using the prediction \citep{Clay75} that excess $^{44}$Ca would exist in the Ti-bearing phases within grains owing to the condensation of radioactive $^{44}$Ti \index{isotopes!44Ti} within a year after the explosion. Because the elemental ratio Ti/Ca is for crystal-lattice reasons larger than solar in SiC grains, the supernova SiC grains possess very large isotopic excess at $^{44}$Ca. These were found in the family of SiC grains called \emph{X grains}, which were already suspected of being supernova grains on the basis of deficiencies in the heavy isotopes of Si and C and on excesses of $^{15}$N \citep{Alex90,Amari92}. \index{isotopes!44Ti} \index{isotopes!17O}

	The extremely large isotopic excesses of $^{26}$Mg and $^{44}$Ca in these SiC X grains were the most dramatic discoveries of fossil extinct radioactivity in stardust. They demonstrated that radioactive $^{26}$Al and $^{44}$Ti had been quite abundant when the grains condensed during the supernova expansion. Predictions had also suggested that fossil abundances in interstellar grains might contribute to excesses of the daughter isotope in solids grown later from interstellar grains (\emph{chemical memory}). But  \citet{Lee77} had emphasized grounds for believing that the excess $^{26}$Mg within solids made in the solar system reflected instead live $^{26}$Al at the time the solids were made. Both fossil and live $^{26}$Al now appear to have existed. These discoveries dramatically spotlighted rich new roles for the astronomy of radioactivity in early solar system chemistry.

	Many other fossil extinct radioactive nuclei have been added to the observed list. In supernova stardust these include excess $^{22}$Ne owing to $^{22}$Na decay, excess $^{41}$K owing to $^{41}$Ca decay, and excess $^{49}$Ti owing to $^{49}$V decay within the grains \citep[see][]{Clay04}. All had been predicted to exist in supernova dust, but their discovery surprised and delighted isotopic chemists and astronomers alike.

	Following the exciting discovery of presolar stardust, its study has emerged as a new area of astronomy. In just 23 years its existence has evolved from a bewildering new discovery into several new techniques for measuring isotopic abundance ratios with high precision in stars. This has been especially important for astronomy with radioactivity, because in stardust the level of extinct radioactivity reflects its abundance during nucleosynthesis in stars, whereas in solar system samples it reflects the level of its survival.  Precisely measured isotopic ratios for four to eight chemical elements endows each gemlike refractory mineral with significance for stellar structure and evolution, and for the chemical evolution of the Milky Way during the epoch 7-5 Gyr ago, and for new insight into nucleosynthesis \citep{Clay04}.  This rich harvest is compromised only by the fact that the donor stars can not be seen, because they died before our solar system began. Their stardust bears no label, save that of their measurable properties. The nature and evolutionary state of the donor stars must be ascertained from the detailed properties, primarily isotopic, of each grain and of the way each grain fits into the spectrum of the thousands of stardust grains that have been analyzed to date. The rapidly growing numbers of analyzed grains allows evolutionary trends within their distinct families to be identified. \footnote{Readers wishing familiarity with these topics can best consult later chapters of this book, or \citet{Clay04}, and {\it Astrophysical Implications of the Laboratory Study of Presolar Materials} \citep{Bern97B}; the {\it Handbook of Isotopes in the Cosmos} by \citet{Clay03} also presents many astrophysical consequences.}

\subsubsection*{Radioactivity and Chemistry of Condensation in Supernovae}
	An unanticipated aspect of astronomy with radioactivity lay in the role of radioactivity in the chemistry of the condensation process. The supernova interior offers a unique laboratory for condensation physics. It guarantees that chemistry begins with gaseous atoms, with no trace of previous molecules or grains. Gamma rays and their Compton-scattered electrons bathe the supernova core. By disrupting the CO molecule, they cause its abundance to be very much smaller than expected from states of chemical thermal equilibrium at the ambient temperature. The small CO molecular abundance enables other non-equilibrium paths to the condensation of carbon-bearing solids. This disequilibrium can be considered to be another aspect of radiogenic luminosity causing a higher degree of excitation in the gas than would be expected in the absence of radioactivity.

        Because the supernova core is hydrogen-free, chemical condensation routes utilizing H are not relevant. Polycyclic aromatic hydrocarbons, for example, do not come into play save in the envelope; but 90-95\% of the mass of ejected Mg, Al and Si emerges in the H-free core. And yet abundant dust condensation is observed to have occurred in SN1987A \citep{Wood93,Col94}, in Cas A \citep{Are99,Dun03} and in the Kepler remnant \citep{Dun03}. \index{supernova!SN1987A} \index{supernova!Cas A} \index{supernova!Kepler} Apparently a few solar masses condensed in Cas A and about one solar mass in Kepler, requiring high condensation efficiency for Mg, Al and Si and even for carbon and thereby establishing supernovae as a major contributor to the budget of thermally condensed interstellar grain cores.

The traditional guideline to condensation of solids had been to assume that chemical equilibrium applies during the expansion and temperature decline of the supernova interior. Formulae yield the equilibrium condensed masses of differing minerals \citep{Latt78,Ebel00}. This approach can not yield grain size but only the total condensed masses. The assumption of thermal equilibrium in the solid phase exaggerates the ability of chemical reactions to maintain chemical equilibrium within the solids as the gas cools. Because of the rapid fall of density and temperature during the expansion of the supernova interior, thermal condensation must be accomplished within about two years, too fast to maintain equilibrium. A nonequilibrium theory of condensation based on nucleation theory followed by subsequent growth has been developed by Kozasa and collaborators \citep{Koz87,Koz89,Koz91,Tod01}. Their method identifies a \emph{key molecular species} whose abundance controls the condensation. However, their questionable assumptions concerning gaseous mixing at the molecular level and on an outdated role for the CO molecule render their results of questionable value.

The nucleosynthesis problems posed by isotopic ratios within individual supernova grains can not be decoupled from physical questions about the chemistry of their condensation. Condensation chemistry is an essential aspect of astronomy with radioactivity. \index{stardust!condensation} Because no single supernova mass zone is able to satisfy the isotopic ratios found in supernova grains,it has long been clear \citep{Amari97} that some type of mixing before condensation is needed to produce their chemical and isotopic compositions; but it is not clear whether that mixing represents atom-scale gaseous mixing in the very young remnant or transport of a growing grain from one composition zone into another. But the biggest question is whether the requirement that the C abundance be greater than O abundance in order to condense SiC and graphite within supernovae is a valid requirement. For their discussion of supernova stardust, for example, \citet{Trav99} took the view that mixing occurs at the atomic level, prior to condensation, and that only C$>$O portions of those mixtures support condensation of carbon. They arbitrarily and instantaneously mixed regions having C$<$O with the He shell having C$>$O to find mixing fractions that retained C$>$O overall. \citet{Koz89} performed similar thought experiments. Such treatment is deeply flawed, however, because it seems certain that gases can not mix at the molecular level within a few years time  \citep[see][]{Den06,Fry91}. The so-called mixing calculated in hydrodynamic studies of supernovae, on the other hand, represents fluids of one composition exchanging places in the homologous expansion with overlying fluids of a different composition. This is a different use of the word \emph{mixing}, and has introduced considerable confusion into published works. Mixing at the molecular level requires very much more time than the year that is available before condensation must occur. Later the expanding remnant is too dilute, its density too small, for condensation.

To circumvent the equilibrium CO trap for carbon,  \citet{Clay99},  \citet{Clay01}, and  \citet{Den06} called upon its dissociation by the fast compton electrons energized by supernova gamma rays. They advanced a kinetic theory of graphite growth and calculated its consequences in detail after advocating a specific nucleation model. The key idea is that small graphite test particles, if placed in a hot gas of C and O atoms, will associate with free C atoms faster than they can be oxidized by free O atoms. This is a property of reaction cross sections rather than of chemical equilibrium. Even though oxidation of carbon would surely be the ultimate end given adequate time, the expansion will terminate the chemistry after about two years with large graphite grains remaining. The graphite essentially is a metastable state of carbon. This theory is supported by observing that supernova 1987A ejected only 10$^{-3}$ $M_{\odot}$ of CO molecules instead of the 0.1 $M_{\odot}$ of CO molecules that is first formed by association reactions in the hot gas before radioactive disruption reduces its abundance \citep{Liu94,Liu95,Gear99}.

	Similar issues probably surround the condensation of supernova SiC. It seems plausible that radioactive liberation of free C atoms from CO molecules will also facilitate the condensation of SiC in O-rich gas; but this is hotly debated. Although a kinetic route to SiC condensation has not been layed out,  \citet{Clay03B}
  \noindent have formulated a  physical description of the ejecta enabling them to make several relevant conclusions derived from assuming that the radioactive CO-disruption mechanism is the correct key to SiC growth. They present animations of a hydrodynamic calculation showing that a reverse shock wave launched toward inner zones by the radially increasing value of the product $\rho r^3$ in the H envelope compresses a dense shell near radial mass coordinate $m=3 M_{\odot}$, where Si and O are the most abundant elements; but some C abundance remains there for possible carbon chemistry. They propose that SiC condenses in that zone, and they detailed several other physical processes that may produce the observed grain compositions. Mixing of a new type during condensation also occurs if the reverse shock generated by the radial expansion colliding with the presupernova wind arrives at the condensation zone at $m = 3 M_{\odot}$ between six months to a year after the explosion, because that reverse shock slows the gas and forces the partially condensed SiC grains to propel forward through the decelerating gas into more $^{29,30}$Si-rich regions, giving perhaps a new interpretation for mixing during condensation.

	Despite many uncertainties, it now appears certain that supernova grains studied by isotopic analysis will provide, through details of condensation chemistry, a new sampling spectrum of young supernova interiors, just as have gamma-ray lines and hard X rays.  If so, a very rich but complex discipline of astronomy with radioactivity will follow.

  \subsubsection*{Astronomy with Radioactivity Today}
	This introductory chapter has focussed on the history and the key physical ideas of astronomy with radioactivity. Effective research depends on a clear grasp of an interdisciplinary set of its central ideas. History itself often provides the best example by which the physical idea can be first grasped. But the remainder of this book looks forward, not backward. Every aspect of astronomy with radioactivity today involves grappling with a host of technical details. This book attempts to bring the reader to that point. Each aspect of astronomy with radioactivity today also involves grappling with the entire world of astronomy. Astronomy itself is many disciplines, and not even astronomers are always comfortable outside their own astronomical technique. How much harder it is to place the complex manifestations of radioactivity into a continuously changing astronomical fabric. But this is the direction of all scientific progress.


%
%
%
%
%
%
%
%
%
\bibliographystyle{spbasic}

\backmatter

\end{document}